\documentclass[aps,prb]{revtex4}
\usepackage{graphicx}
\usepackage{psfrag}
\def\figtag{Fig.}
\def\reftag{Ref.}
\def\eqtag{Eq.}
\def\tabletag{Table }
\def\gsim{\buildrel {\textstyle >}\over {_\sim}}
\def\lsim{\buildrel {\textstyle <}\over {_\sim}}
\def\figstag{Figs.}
\def\refstag{Refs.}

\begin{document}

\title{Ordering of the Heisenberg spin glass in high
dimensions}
\author{Daisuke Imagawa}
\email{imag@spin.ess.sci.osaka-u.ac.jp}

\author{Hikaru Kawamura}
\email{kawamura@ess.sci.osaka-u.ac.jp}
\homepage{http://thmat8.ess.sci.osaka-u.ac.jp/~kawamura}

\affiliation{Graduate School of Science, Osaka University,
Toyonaka, Osaka 560-0043, Japan}

\date{\today}

\begin{abstract}
Ordering of the Heisenberg spin glass with the nearest-neighbor
Gaussian coupling is investigated by equilibrium Monte Carlo simulations
in four and five dimensions. Ordering
of the mean-field Heisenberg spin glass
is also studied for comparison.
Particular attention is paid to the nature of the
spin-glass and chiral-glass orderings.
Our numerical data suggest that, in five dimensions,
the model
exhibits a single spin-glass transition at a finite temperature,
where the spin-glass order accompanying the simultaneous
chiral-glass order sets in.
In four dimensions, the
model exhibits a marginal behavior.
Chiral-glass transition at a finite temperature not
accompanying the standard spin-glass order is likely to occur, while
the critical region associated with the chiral-glass
transition is very narrow suggesting that the dimension four
is close to the marginal dimensionality.
\end{abstract}
\maketitle

\section{Introduction}

In numerical studies of spin glasses (SGs), much effort has
been devoted to clarify the properties of the so-called
Edwards-Anderson (EA) model~\cite{SGrev}.
Most of these numerical works on the EA model have concentrated
on the
Ising
EA model. It is also very important, however,
to clarify
the properties of the corresponding
Heisenberg
model.
This is simply due to the
fact that many of real
SG magnets are Heisenberg-like rather than Ising-like in the sense that
the magnetic anisotropy is considerably weaker than the isotropic
exchange interaction.~\cite{SGrev,OYS}

Indeed, several numerical works have been performed
on the Heisenberg EA model.
Earlier numerical studies suggested
that, in apparent contrast to experiments, the isotropic Heisenberg SG
in three dimensions (3D) did not exhibit an equilibrium SG transition
at any finite temperature.
~\cite{OYS,Banavar,McMillan,Matsubara1,Kawamura92,Kawamura95,Yoshino}
This observation
leads to general belief that the weak
random magnetic anisotropy is crucially important in realizing
a finite-temperature SG transition and a stable SG phase,
which causes a crossover from the $T_g=0$ isotropic
Heisenberg behavior to the $T_g>0$ anisotropic Ising behavior.
The expected Heisenberg-to-Ising
crossover, however, has not been observed experimentally,
and this puzzle has remained unexplained.~\cite{SGrev,OYS}

Meanwhile, a novel
possibility was suggested by one of the present authors
(H.K.)
that the 3D Heisenberg SG might exhibit an equilibrium phase
transition at a finite temperature,
not in the spin sector as usually envisaged,
but in the 
chirality
sector,
i.e.,
might exhibit a 
chiral-glass
transition.~\cite{Kawamura92}
Chirality is a multispin variable
representing the sense or the handedness of local
noncoplanar spin structures induced by spin frustration.
In the chiral-glass ordered
state, the chirality is ordered in a spatially random manner while
the Heisenberg spin remains paramagnetic. 
References ~\cite{Kawamura92,Kawamura95,
Kawamura98,HK1,3DHSGinH_1,3DHSGinH_2} claimed that
the standard SG order
associated with the freezing of the Heisenberg spin occurred at a temperature
lower than the chiral-glass transition temperature at $T=T_{\rm SG}<
T_{\rm CG}$, quite possibly $T_{\rm SG}=0$.
It means that the spin and the
chirality are decoupled on long length scales
(spin-chirality decoupling). In fact,
based on such a spin-chirality decoupling
picture, a chirality scenario of the SG transition
has been advanced,
which  explains the experimentally observed SG transition as essentially
chirality driven~\cite{Kawamura92,Kawamura98}.
Note that the numerical observation
of a finite-temperature chiral-glass transition in the 3D Heisenberg SG of
\refstag~\cite{Kawamura92, Kawamura95, Kawamura98, HK1,3DHSGinH_1,3DHSGinH_2}
is not inconsistent with the earlier observations
of the absence of the conventional SG order at any finite temperature.

Recently, however, in a series of numerical studies on the 3D
Heisenberg EA model, Tohoku group criticized the earlier numerical works,
claiming that in the 3D Heisenberg SG the spin ordered at a finite temperature
and that the SG transition temperature
might coincide with the chiral-glass transition temperature,
{\it i.e.\/}, $T_{{\rm SG}}=T_{{\rm CG}}>0$~\cite{Matsubara2,Nakamura}.
By contrast, Hukushima and Kawamura maintained that
in 3D the spin and the
chirality were decoupled on sufficiently long length scales, and that
$T_{{\rm SG}}<T_{{\rm CG}}$~\cite{HK2}, supporting the earlier numerical
results. The situation in 3D thus remains controversial.

Under such circumstances,
in order to shed further light on the nature of the
ordering in 3D,
it might be useful to study the
problem for the general space dimensionality $D$, particularly for
dimensions higher than $D=3$. In the limit of
infinite dimensions $D\rightarrow \infty$, the model
reduces to the corresponding mean-field model, {\it i.e.\/},
the Heisenberg Sherrington-Kirkpatrick (SK)
model. In the case of equal weights of the
ferromagnetic and antiferromagnetic
interactions,
the SK model is known to exhibit a single continuous SG transition.
Hence, in the $D\rightarrow \infty$ limit,
the order parameter of the transition is the Heisenberg spin itself, with no
exotic phase such as the chiral-glass phase.
Furthermore, the SG ordered state of the SK model is known to
exhibit a hierarchical type of replica-symmetry breaking (RSB),
{\it i.e.\/}, a full RSB.

Then, questions which naturally arise are:
(i) What is the lower critical dimension
(LCD) of
the SG order $d_\ell^{{\rm SG}}$? (ii) Is $d_\ell^{{\rm SG}}$ the same as
the LCD associated with the chiral-glass order $d_\ell^{{\rm CG}}$?

Concerning the point (i), several earlier
numerical studies including the high-temperature expansion~\cite{HighT_R}
and the numerical domain-wall renormalization-group calculation
~\cite{Cieplak, McMillan}
suggested that $d_\ell^{{\rm SG}}$
might be close to four.
Meanwhile, Anderson and Pond argued that $d_\ell^{{\rm SG}}=3$~\cite{Anderson}.
First Monte Carlo (MC)
simulation on the high-dimensional Heisenberg EA model was
performed by Stauffer and Binder~\cite{Stauffer}.
By studying the temporal decay of the EA order parameter,
they suggested that a finite-temperature SG order
occurred in $D=5$ and $6$, but not in $D\leq 4$~\cite{Stauffer}.
More recently, the 4D Heisenberg EA model
was studied by Coluzzi by equilibrium
MC simulation~\cite{Coluzzi}.
By examining the behavior of the Binder ratio,
she suggested the occurrence of a finite-temperature SG transition
in $D=4$ in contrast to the suggestion of \reftag~\cite{Stauffer}.
There seems to be no consensus as to the point (i).

The point (ii) above is closely related to the controversy regarding
whether the spin-glass and the chiral-glass orders occur simultaneously or
separately in 3D. To the authors' knowledge, concerning the chiral-glass order
in $D\geq 4$ dimensions,
no calculation has been reported so far.
In the present paper, we wish to fill this gap. We study both the
spin-glass and the chiral-glass orders of the Heisenberg EA model
in both 4D and 5D by means of a large-scale equilibrium MC
simulation.
In particular, we simulate larger lattices and lower temperatures
than those covered in \reftag~\cite{Coluzzi}.
For comparison, a simulation is also performed on the
mean-field Heisenberg SK model corresponding to $D=\infty$.

Our data suggest that, in 5D, the model
exhibits a single SG transition at a finite temperature
reminiscent to the one of the Heisenberg SK model.
The chirality orders simultaneously with the spin, but it behaves as the
composite operator of the spin, not as the order parameter.
The SG ordered state in 5D accompanies a peculiar type of RSB,
most probably a one-step-like RSB, which is different in character
from the full RSB realized in the SK model.
In 4D, by contrast, the
model exhibits a pure chiral-glass transition at a finite temperature, not
accompanying the standard SG order.
The critical region associated with the chiral-glass
transition, however, is very narrow, suggesting that the 4D
model lies close to the marginal dimensionality.
The chiral-glass ordered state accompanies a one-step-like RSB.

The present paper is organized as follows.
In \S\ref{secModel}, we introduce our model and explain some of the
details of the MC calculation. Various physical
quantities calculated in our MC simulation are defined in \S\ref{secPhysQ}.
The results of our MC simulation
on the 4D, 5D and SK models are presented in \S\ref{secResult}.
Section \ref{Summary} is devoted to summary and discussion.

\section{The model and the method}
\label{secModel}

The model we consider is the isotropic classical Heisenberg
model on a 4D or 5D hypercubic lattice,
with the nearest-neighbor Gaussian coupling.
The Hamiltonian is given by
\begin{equation}
{\cal H}=-\sum_{<ij>}J_{ij}\vec{S}_i\cdot \vec{S}_j\ \ ,
\label{eqn:hamil}
\end{equation}
where $\vec{S}_i=(S_i^x,S_i^y,S_i^z)$ is a three-component unit vector,
and $<ij>$ sum is taken over nearest-neighbor pairs on the lattice.
The nearest-neighbor coupling $J_{ij}$ is assumed to
obey the Gaussian distribution with a zero mean and a variance $J^2$.

For comparison, we also simulate the corresponding infinite-ranged model,
{\it i.e.\/}, the Heisenberg SK model
corresponding to $D\rightarrow \infty$.
In the SK model, the Gaussian coupling $J_{ij}$ works between all possible
pairs of total $N$ spins with a zero mean and a variance
$J^{2}/N$.

We perform an equilibrium MC simulation on these models.
In 4D, the lattices studied are the hypercubic lattices with $N=L^{4}$ sites
with $L=4$, 6, 8, and 10, whereas in 5D,
$N=L^{5}$ with $L=3$, 4, 5, 6 and 7. In the case of the SK model,
$N$ is taken to be $N=32$, 64, 128, 256 and 512.
In all cases, we impose periodic boundary conditions in all $D$ directions.
The sample average is taken over 64-200 independent bond realizations,
depending on the system size $L$ and the lattice dimensionality $D$.
Error bars of
physical quantities are estimated by the sample-to-sample statistical
fluctuation over the bond realizations.

In order to facilitate efficient thermalization, we combine the standard
heat-bath method with the temperature-exchange technique~\cite{TempExMC}.
Care is taken to be sure that
the system is fully equilibrated.
Equilibration is checked by the following procedures.
First, we monitor the system to travel back and forth
many times along the
temperature axis during the
the temperature-exchange process (typically more than 10 times)
between the maximum and minimum temperature points.
We check at the same time
that the relaxation
due to the standard heat-bath updating
is reasonably fast at the highest temperature,
whose relaxation time is of order $10^2$ Monte Carlo steps
per spin (MCS). This guarantees that different parts of
the phase space are sampled in each ``cycle'' of the temperature-exchange
run. Second, we check
the stability of the results against at least three times longer runs
for a subset of samples.
Third, we use the method recently developed in
\refstag~\cite{3DISG_Y,4DXY} for the Gaussian coupling, in which a certain quantity
is calculated in two ways, each of which is expected to approach the
asymptotic equilibrium value either from above or from below.
Further details of our MC simulation
are given in \tabletag\ref{table-MCparam}.
%
%
\begin{table}
\begin{center}
\caption{Detailed conditions of the MC simulation. Here,
$D$ represents the spatial dimensionality,
$N$ the total number of spins,
$N_{\rm samp}$ the total number of samples, $N_{\rm T}$ the total number of
temperature points used in the temperature-exchange run,
$T_{\max}/J$ and $T_{\min}/J$ the maximum and minimum
temperatures in the temperature-exchange run.}
\label{table-MCparam}
\scalebox{0.5}{\includegraphics[width=\columnwidth]{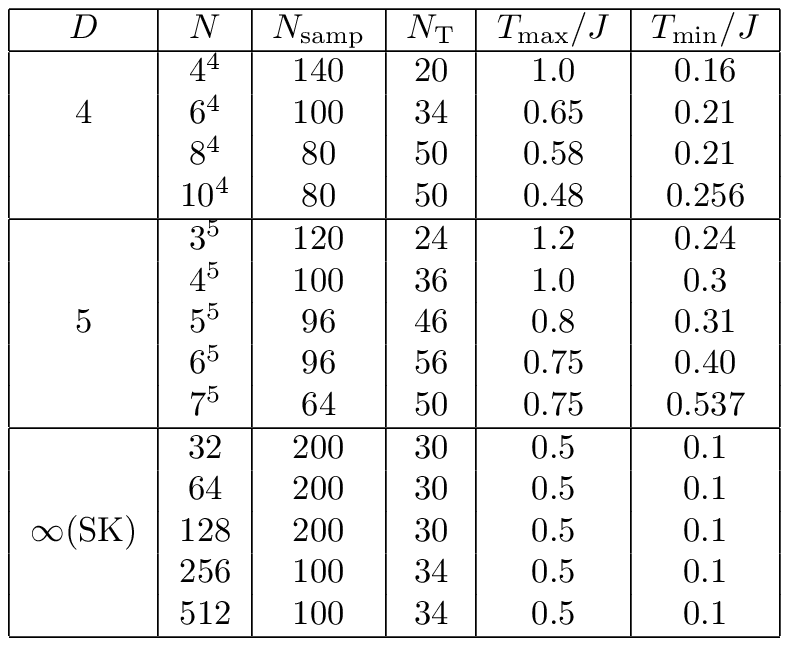}}
\end{center}
\end{table}

%
\section{Physical Quantities}
\label{secPhysQ}
In this section, we
define various physical quantities calculated in our
simulation below.

\subsection{Spin-related quantities}

By considering two independent systems (``replicas'') described by
the same Hamiltonian (\ref{eqn:hamil}),
one can define an overlap variable.
The overlap of the Heisenberg spin
is defined
as a {\it tensor\/} variable $q_{\mu\nu}$
between the $\mu$ and $\nu$
components ($\mu$, $\nu$=$x,y,z$) of the Heisenberg spin,
\begin{equation}
q_{\mu\nu}=\frac{1}{N}\sum_{i=1}^N S_{i\mu}^{(1)}S_{i\nu}^{(2)}\ \ ,
\ \ (\mu,\nu=x,y,z)\ \ ,
\end{equation}
where $\vec{S}_i^{(1)}$ and $\vec{S}_i^{(2)}$ are the $i$-th
Heisenberg spins of the replicas 1 and 2, respectively.
In our simulation, we prepare the two replicas 1 and 2 by
running two independent sequences of systems
in parallel with different spin initial conditions and
different sequences of random numbers.
In terms of these tensor overlaps, the SG order parameter is defined by
\begin{equation}
q_{\rm s}^{(2)} = [\langle q_{\rm s}^2\rangle]\ \ ,
\ \ \ \ q_{\rm s}^2 = \sum_{\mu,\nu=x,y,z}q_{\mu\nu}^2\ \ ,
\end{equation}
while the associated spin Binder
ratio is defined by
\begin{equation}
g_{\rm s} = \frac{1}{2}
\left(11 - 9\frac{[\langle q_{\rm s}^4\rangle]}
{[\langle q_{\rm s}^2\rangle]^2}\right)\ \ ,
\label{eqn:gs_def}
\end{equation}
where $\langle\cdots\rangle$ represents the thermal average and
[$\cdots$] the average over the bond disorder.
Note that $g_{\rm s}$ is
normalized here so that, in the thermodynamic limit,
they vanish in the high-temperature phase and gives unity in the
nondegenrate ordered state.

The spin-overlap distribution function is originally defined in the
tensor space, since the relevant spin-overlap
has $3\times 3=9$ independent components.
For the convenience of illustration,
we introduce here the diagonal spin-overlap distribution function,
\begin{equation}
P_{\rm s}(q_{\rm diag}^{\prime})=[\langle\delta
(q_{\rm diag}^{\prime}-q_{\rm diag})\rangle]\ \ ,
\end{equation}
defined in terms of the diagonal
overlap $q_{\rm diag}$ which is the trace of the tensor overlap
$q_{\mu\nu}$'s,
\begin{equation}
q_{{\rm diag}}=\sum _{\mu=x,y,z} q_{\mu \mu}
       =\frac{1}{N}\sum_{i=1}^N \vec{S}_{i}^{(1)}\cdot\vec{S}_{i}^{(2)}\ \ .
\label{eqn:qd}
\end{equation}
Note that the spin Binder ratio \eqtag(\ref{eqn:gs_def}) is
defined from the full tensor overlap distribution function,
but cannot be derived solely from the diagonal overlap
distribution function \eqtag(\ref{eqn:qd}).

In zero field,
the distribution function $P_{\rm s}(q_{\rm diag})$ is symmetric
with respect to $q_{\rm diag}=0$.
In the high-temperature phase,
each $q_{\mu\nu}$ ($\mu, \nu=x,y,z$) is expected
to be Gaussian-distributed around
$q_{\mu\nu}=0$ in the $L\rightarrow \infty$ limit, and so is
$q_{\rm diag}$.
In the hypothetical SG ordered state, reflecting the fact that $q_{\rm diag}$ transforms nontrivially under
the independent global $O(3)$ spin rotations on the two replicas, even a
self-overlap part of $P_{\rm s}(q_{\rm diag})$ develops a nontrivial
shape, {\it i,e,\/}, it is not a simple delta-function
located at the EA SG order parameter
$\pm q_{\rm s}^{{\rm EA}}$. Hence, we need to clarify
first how the function $P_{\rm s}(q_{\rm diag})$ looks
like in the possible SG ordered state
with a nonzero $\pm q_{\rm s}^{{\rm EA}}$~\cite{3DXY}.

Let us hypothesize here that there exists a
{\it spin\/}-glass ordered state
characterized by a nonzero EA SG order parameter
$q_{\rm s}^{\rm EA}>0$.
One can show in the $L\rightarrow \infty$ limit that
the self-overlap part of $P_{\rm s}(q_{\rm diag})$ is given by
\begin{equation}
P_{\rm s}(q_{{\rm diag}})=
\frac{3\sqrt{3}}{4\pi q_{\rm s}^{\rm EA}}\left(
\sqrt{\frac{q_{\rm s}^{\rm EA}-q_{\rm diag}}
{3q_{\rm diag}+q_{\rm s}^{\rm EA}}}
+\sqrt{\frac{q_{\rm s}^{\rm EA}+q_{\rm diag}}
{-3q_{\rm diag}+q_{\rm s}^{\rm EA}}}
\right)\ \ ,
\label{eqn:Pdform}
\end{equation}
which is illustrated in \figtag\ref{fig_LRO}.
The derivation of \eqtag(\ref{eqn:Pdform}) is given in the
Appendix.
Note that diverging $\delta$-function peaks
appear at
$q_{\rm diag}=\pm \frac{1}{3}q_{\rm s}^{\rm EA}$,
not at $q_{\rm diag}=\pm q_{\rm s}^{\rm EA}$.
If the SG ordered state accompanies RSB,
the associated nontrivial contribution would be added to
the one given by \eqtag(\ref{eqn:Pdform}).
In any case, an important observation
here is that, as long as the ordered state possesses a
finite SG long-range order (LRO),
the diverging peak should arise in $P_{\rm s}(q_{{\rm diag}})$
at $q_{{\rm diag}}=\pm \frac{1}{3}q_{\rm s}^{\rm EA}$.
\begin{figure}[ht]
\begin{center}
\includegraphics[scale=0.7]{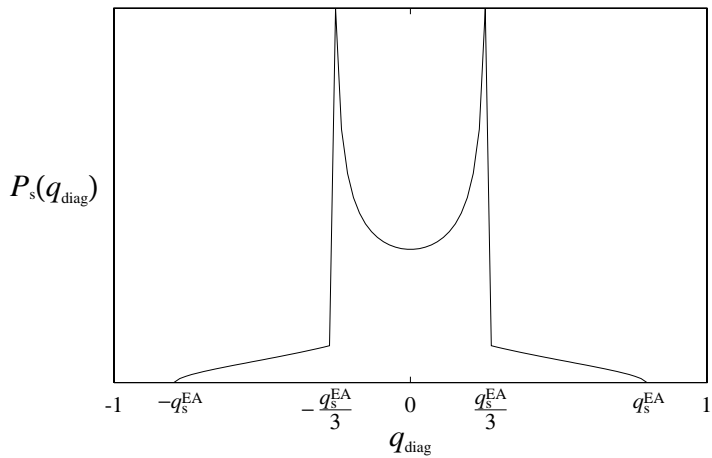}
\caption{Sketch of the self-overlap part of the
diagonal spin-overlap distribution function
$P_{\rm s}(q_{{\rm diag}})$
in the thermodynamic limit,
expected when there
exists a finite SG long-range order with a nonzero
$q_{\rm s}^{{\rm EA}}>0$.}
\label{fig_LRO}
\end{center}
\end{figure}

%
\subsection{Chirality-related quantities}

We define
the local chirality at the $i$-th site and in the $\mu$-th
direction, $\chi_{i\mu}$, for three neighboring Heisenberg spins
by the scalar
\begin{equation}
\chi_{i\mu}=
\vec{S}_{i+{\hat{e}}_{\mu}}\cdot
(\vec{S}_i\times\vec{S}_{i-{\hat{e}}_{\mu}})\ \ ,
\end{equation}
where ${\hat{e}}_{\mu}\ (\mu=x_1,x_2,\cdots,x_D)$ denotes
a unit vector along the $\mu$-th axis.
By this definition, there are in total $DN$ local
chiral variables.

In the case of the SK model, since
there is no ``neighbors'' in the model, the definition
of the local chirality accompanies some difficulties.
Here, just for convenience, we number the $N$ spins
arbitrarily, and
define the local chirality by
\begin{equation}
\chi_{i}=\vec{S}_{i+1}\cdot(\vec{S}_i\times\vec{S}_{i-1})\ \ .
\end{equation}
Then, there are in total $N$ chiral variables.

The mean local amplitude of the chirality, $\bar \chi$, may be defined by
\begin{equation}
\bar{\chi}=\sqrt{\frac{1}{DN}\sum_{i=1}^N
\sum_{\mu=x_1,x_2,\cdots,x_D}[\langle\chi_{i\mu}^2\rangle]}\ \ ,
\end{equation}
for $D=4$ and $5$, and by
\begin{equation}
\bar{\chi}=\sqrt{\frac{1}{N}\sum_{i=1}^N[\langle\chi_{i}^2\rangle]}\ \ ,
\end{equation}
for the SK model. Note that the magnitude of $\bar{\chi}$
tells us the extent of the noncoplanarity of the local spin
structures. In particular,
this quantity vanishes for any coplanar spin configuration.

As in the case of the Heisenberg spin,
one can define an overlap of the chiral variable
by considering the two replicas. In the cases of $D=4$ and
$5$, it is defined by
\begin{equation}
q_{\chi}=
\frac{1}{DN}\sum_{i=1}^N\sum_{\mu=x_1,x_2,\cdots,x_D}
\chi_{i\mu}^{(1)}\chi_{i\mu}^{(2)}\ \ ,
\end{equation}
where $\chi_{i\mu}^{(1)}$ and $\chi_{i\mu}^{(2)}$ represent the chiral
variables of the replicas 1 and 2, respectively. In the case of the
SK model, it is defined by
\begin{equation}
q_{\chi}=
\frac{1}{N}\sum_{i=1}^N\chi_{i}^{(1)}\chi_{i}^{(2)}\ \ .
\end{equation}
In terms of this chiral overlap $q_{\chi}$, the chiral-glass
order parameter is defined by
\begin{equation}
q_{\chi}^{(2)}=[\langle q_{\chi}^2\rangle]\ \ .
\end{equation}
The associated chiral-glass susceptibility may be defined by
\begin{equation}
\chi_\chi=DN[\langle q_{\chi}^2\rangle]\ \ ,
\end{equation}
in the cases of $D=4$ and $5$,
while in the case of the SK model, it is defined by
\begin{equation}
\chi_\chi=N[\langle q_{\chi}^2\rangle]\ \ .
\end{equation}
Unlike the spin variable, the local magnitude
of the chirality is somewhat temperature dependent.
In order to take account of this effect,
we also consider the reduced chiral-glass susceptibility
$\tilde \chi_\chi$ by dividing $\chi_{\chi}$ by the appropriate powers of $\bar \chi$,
\begin{equation}
\tilde \chi_\chi=\frac{\chi_\chi}{\bar \chi^4}\ \ .
\end{equation}
The Binder ratio of the chirality is defined by
\begin{equation}
g_{\chi}=
\frac{1}{2}
\left(3-\frac{[\langle q_{\chi}^4\rangle]}
{[\langle q_{\chi}^2\rangle]^2}\right)\ \ .
\end{equation}
The distribution function of the chiral overlap
$q_{\chi}$ is defined by
\begin{equation}
P_\chi(q^{\prime}_{\chi})=[\langle\delta(q_\chi^{\prime}-q_{\chi})\rangle]\ \ .
\end{equation}
%

\section{Monte Carlo Results}
\label{secResult}

In this section, we present our MC results on the Heisenberg
EA models in 4D, 5D, and in the SK limit (corresponding to $D=\infty$).
\subsection{Chiral Binder ratio}

\begin{figure}[ht]
\leavevmode
\begin{center}
\begin{tabular}{l}
\includegraphics[scale=0.7]{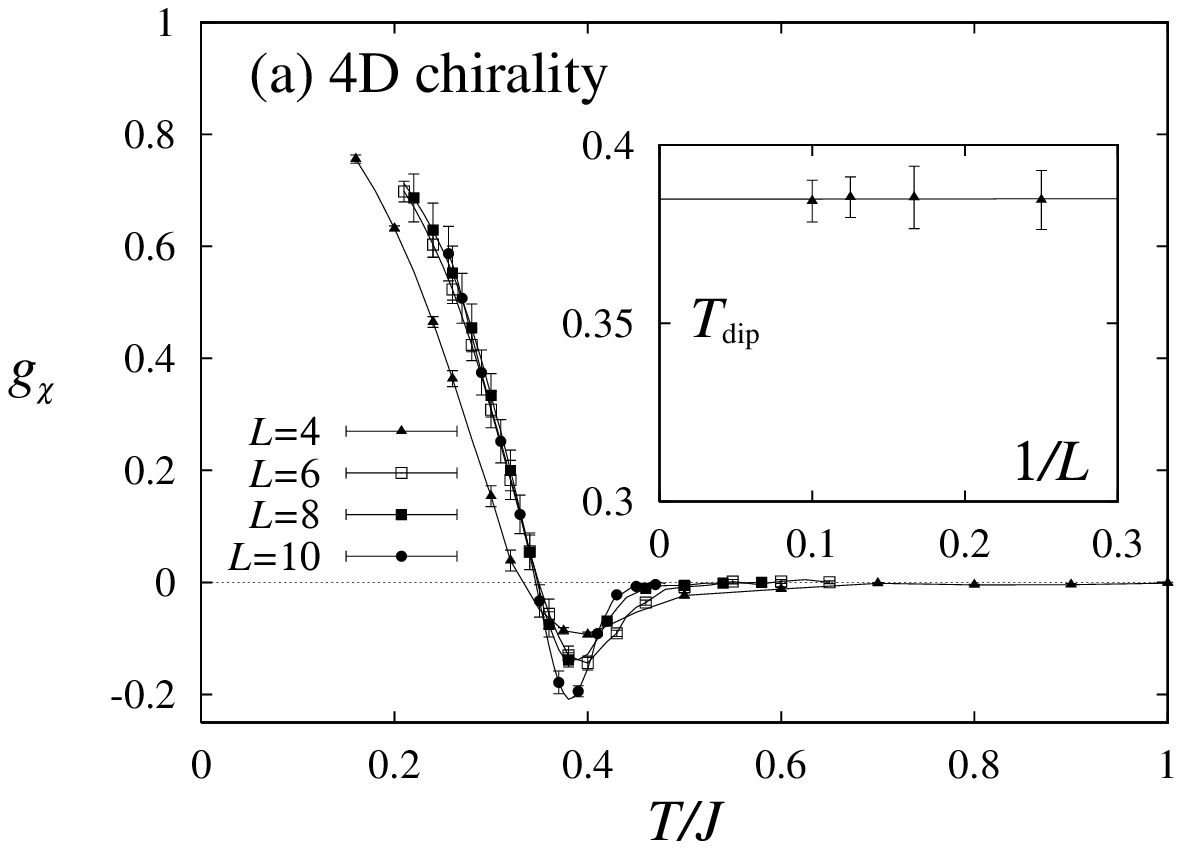}
\end{tabular}
\begin{tabular}{ll}
\includegraphics[scale=0.7]{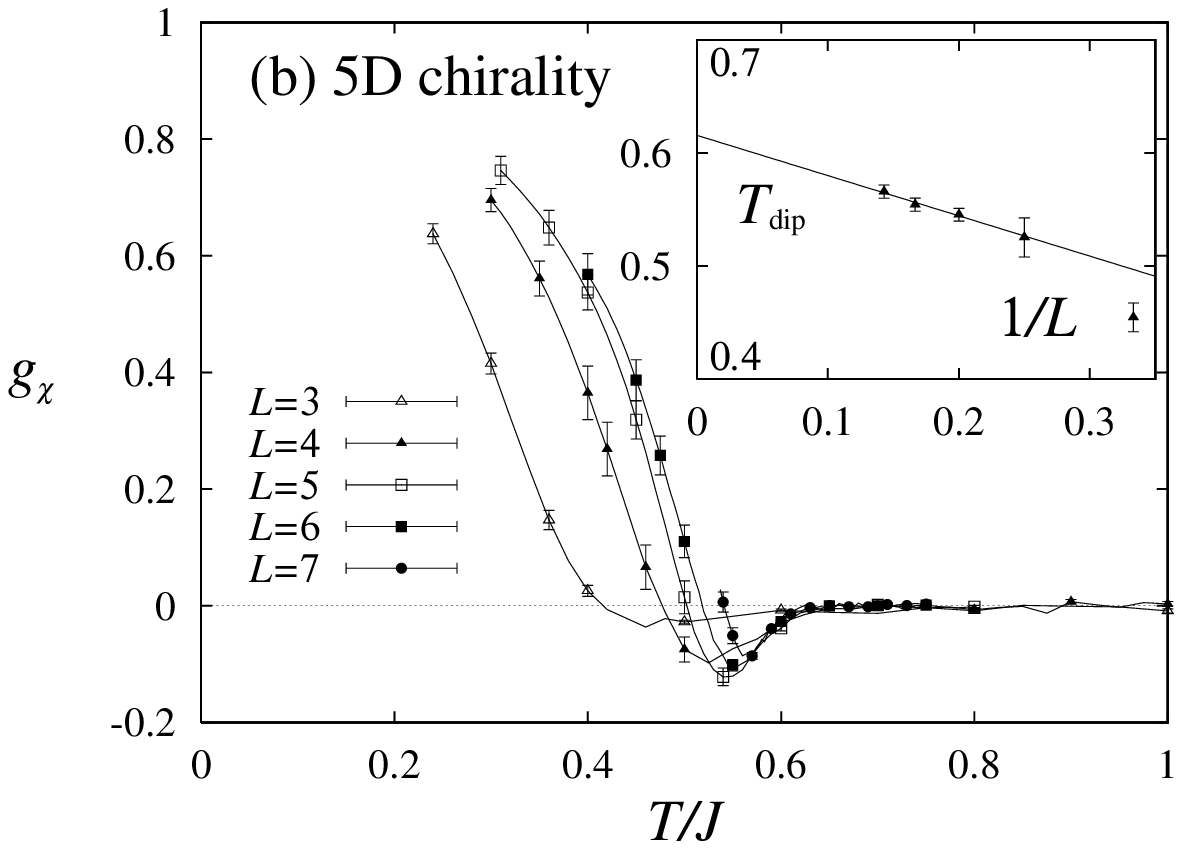}
&
\includegraphics[scale=0.7]{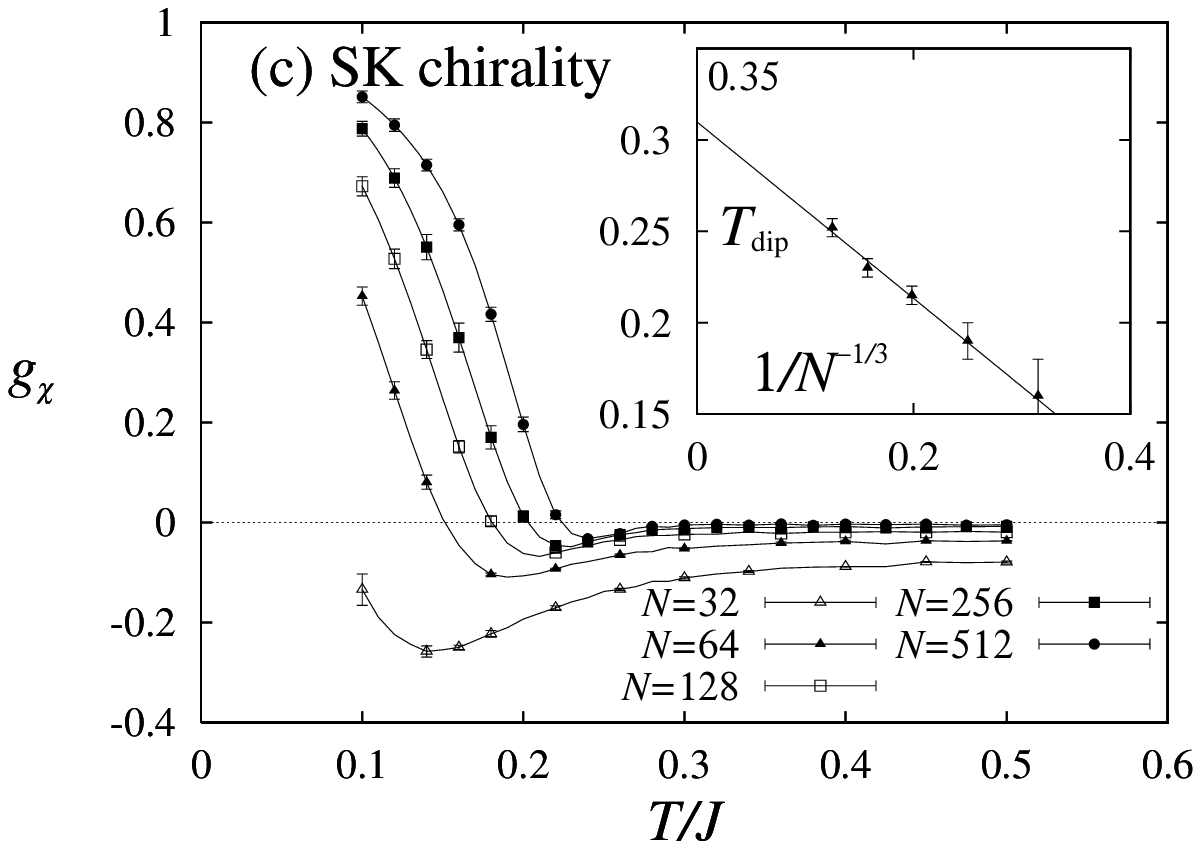}
\end{tabular}
\caption{Temperature and size dependence of the chiral Binder ratio $g_{\chi}$
in 4D (a), in 5D (b), and in the SK model (c). In the insets,
$T_{\rm dip}$ is plotted versus $1/L$ (or versus $1/N^{-1/3}$
in the case of the SK model).
}
\label{fig_gx}
\end{center}
\end{figure}

In \figtag\ref{fig_gx}, we show the Binder ratio of the chirality in the
cases of
(a) 4D, (b) 5D and (c) the SK model, respectively.
In all these cases, $g_\chi$ exhibits a negative dip, while
its temperature and size dependence is somewhat different from each other.

In 4D, with increasing the lattice size $L$, the negative dip tends to deepen
while the dip temperature $T_{\rm dip}$ is almost kept constant at
around $T/J=0.38$: See the inset of \figtag\ref{fig_gx}.
One can also see from \figtag\ref{fig_gx} that
$g_{\chi}$ for various $L$ cross in the negative region of $g_{\chi}$
at temperatures slightly above $T_{\rm dip}$.
The occurrence of a negative dip deepening with $L$, accompanied by a
crossing on the negative side of $g_{\chi}$,
is similar to the one previously observed in the corresponding
3D model~\cite{HK1,3DHSGinH_1,3DHSGinH_2},
although, in 3D, $T_{\rm dip}$ tends to shift toward lower temperature
with increasing $L$.
As argued in \reftag~\cite{HK1,3DHSGinH_1,3DHSGinH_2},
the occurrence of a negative
dip deepening with $L$ is a signature of the occurrence
of a phase transition in the chiral sector.
By making a linear extrapolation of $T_{\rm dip}(L)$ with
respect to $L^{-1}$, as shown in the inset of \figtag\ref{fig_gx} (a),
we  estimate the bulk
chiral-glass transition temperature
as $T_{\rm CG}/J=0.38(2)$.
Below $T_{{\rm CG}}$, the
curves for $L\geq 6$ almost merge into a curve exhibiting the nontrivial
temperature dependence. Such a behavior suggests that the chiral
ordered state accompanies
a nontrivial phase-space structure, {\it i.e.\/}, RSB.

In 5D,
although the negative dip tends to deepen up to the size $L=5$,
it tends to become shallower for $L\geq 6$.
In contrast to the 3D and 4D cases,
the dip temperature $T_{\rm dip}$ tends to shift
toward higher temperature with increasing $L$.
The observed temperature and size dependence of $g_\chi$ strongly
suggests again that the limit $T_{\rm dip}(L\rightarrow \infty)$
corresponds to a transition temperature in the chiral sector.
The $1/L$-extrapolation of $T_{\rm dip}(L)$ to $L\rightarrow \infty$,
shown in the inset of \figtag\ref{fig_gx} (b), yields the estimate
$T_{\rm CG}/J=0.62(2)$.

In the SK case,
the negative dip of $g_{\chi}$ becomes shallower with increasing $L$, and
$T_{\rm dip}$ shifts
toward higher temperature. In the inset, $T_{\rm dip}(N)$ is plotted
as a function of $N^{-1/3}$~\cite{InfSys_FSS,KH_RSB}, which yields
$T_{\rm CG}/J=0.31(2)$.
The estimated chiral-glass ordering
temperature agrees within errors with the exactly-known
SG transition temperature of
the Heisenberg SK model, $T_{{\rm SG}}/J=1/3$. This coincidence simply
confirms the fact that, at the SG transition of the Heisenberg SK model,
Heisenberg spins order in a noncoplanar manner, which necessarily
accompanies the onset of a nonzero chiral-glass LRO.
As is evident, in the SK case,
the order parameter of the transition is not the chirality, but the
Heisenberg spin itself.

\subsection{Chiral autocorrelation function}
More direct measure of the chiral-glass transition
may be obtained from
the equilibrium dynamics of the model.
We compute the autocorrelation function of the chirality defined by
\begin{equation}
C_{\chi}(t)=\frac{1}{DN}\sum_{i=1}^N\sum_{\mu=x_1,x_2,\cdots,x_D}
[\langle\chi_{i\mu}(t_0)\chi_{i\mu}(t+t_0)\rangle ]\ \ ,
\label{Cxt}
\end{equation}
where the ``time'' $t$ is measured here in units of MCS.
In computing (\ref{Cxt}), the simulation is performed
according to the standard heat-bath updating without the
temperature-exchange procedure, while
the starting spin configuration at $t=t_0$ is taken from
the equilibrium spin configurations
generated in our temperature-exchange MC runs.

\begin{figure}[ht]
\leavevmode
\begin{center}
\begin{tabular}{ll}
\includegraphics[scale=0.7]{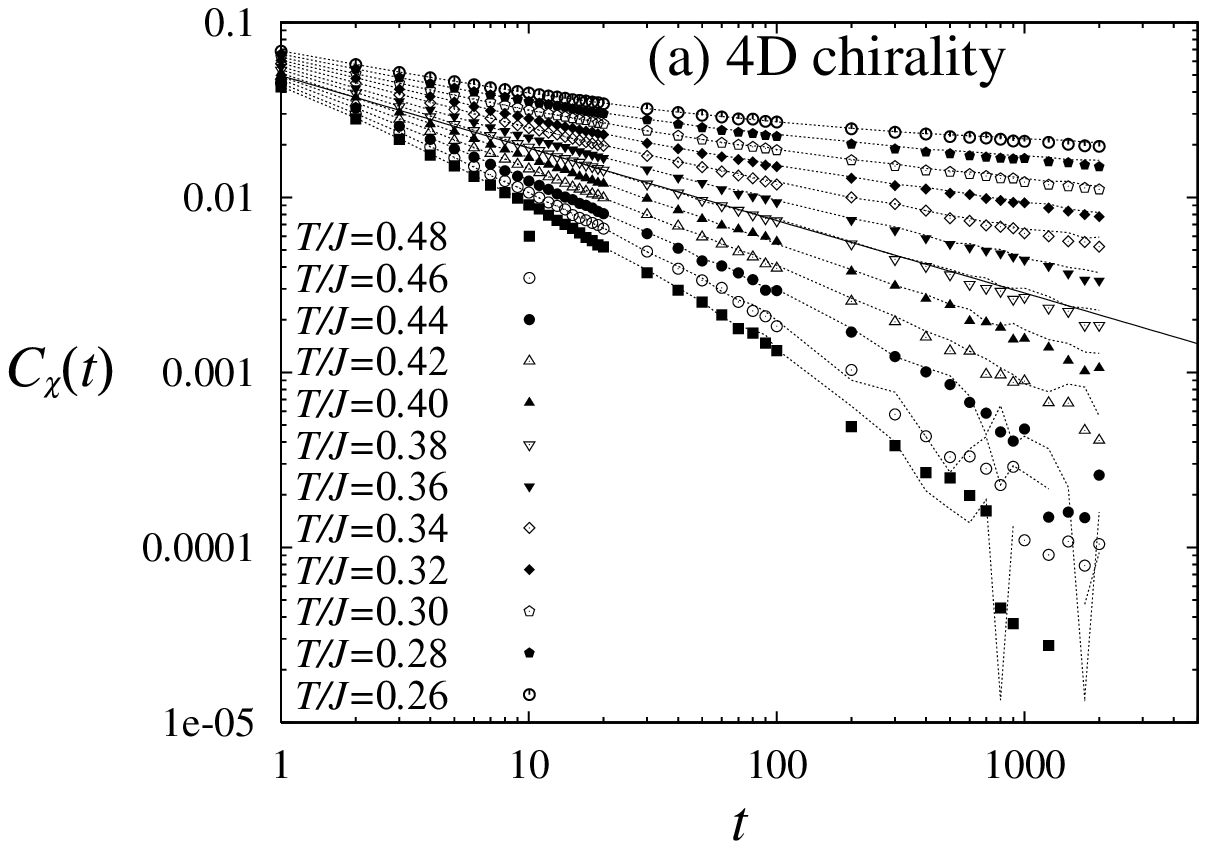}
&
\includegraphics[scale=0.7]{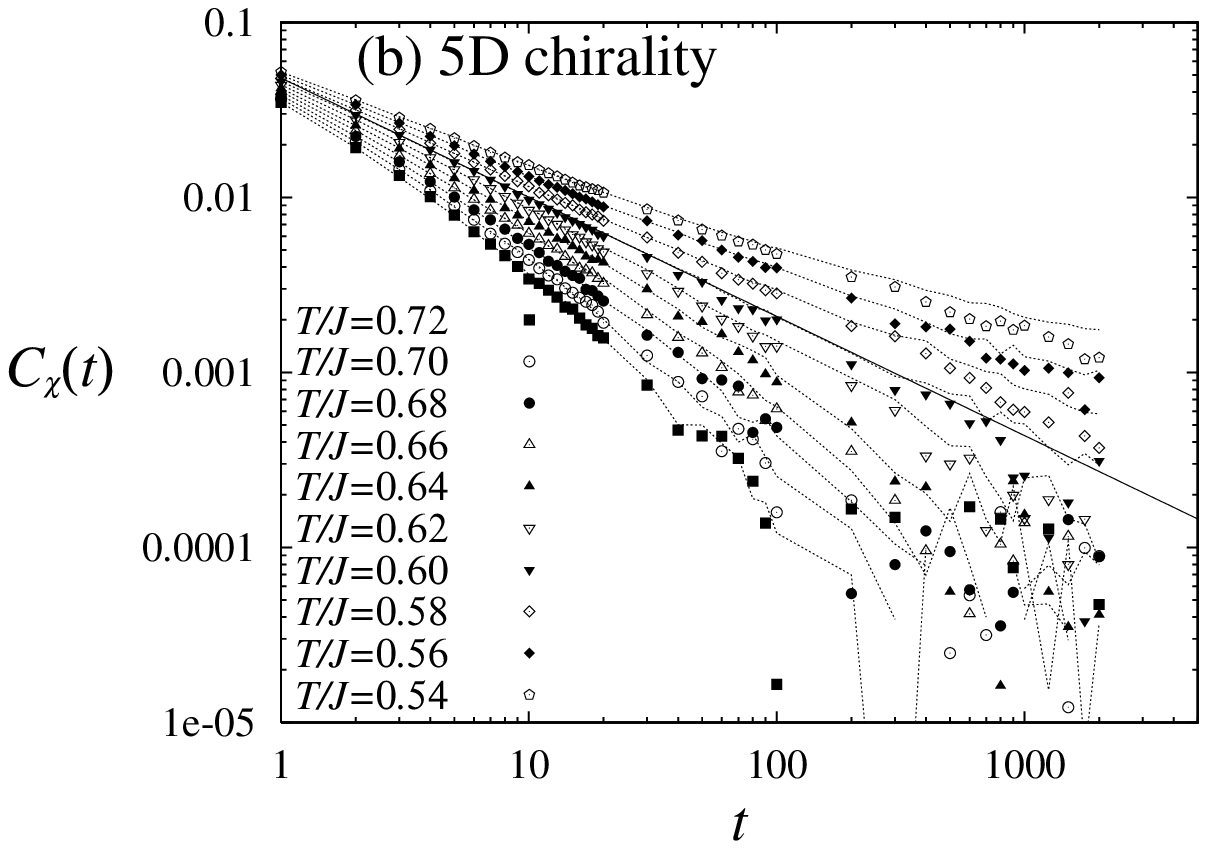}
\end{tabular}
\caption{Temporal decay of the equilibrium
chiral autocorrelation function at various temperatures
in  4D (a), and in 5D (b). In (a),
symbols represent the data of $L=8$ and thin lines those of $L=10$.
In (b), symbols represent the data of $L=6$ and thin lines those of $L=7$.
Solid straight lines represent the power-law fits of the data
at $T/J=0.38$  (a), and at $T/J=0.60$ (b).
}
\label{fig_cx}
\end{center}
\end{figure}

We show in \figtag\ref{fig_cx} the time dependence of
$C_{\chi}(t)$ on a log-log plot
for the cases of (a) 4D, and (b) 5D. To check the possible size dependence,
the data for the two largest lattice sizes are given
together, one denoted by symbols and the other by thin lines.
In the chiral-glass ordered state with a nonzero $q_\chi^{{\rm EA}}$,
$C_{\chi}(t)$ in the $L\rightarrow \infty$ limit should exhibit
an upward curvature, tending to
$q_\chi^{\rm EA}>0$.
In the disordered phase, $C_{\chi}(t)$
should exhibit a downward curvature.
Just at $T=T_{\rm CG}$, the linear behavior corresponding to
the power-law decay is expected.
As shown in the figures, in the time region
where the finite-size effect is negligible,
$C_{\chi}(t)$ shows either a downward curvature
characteristic of the disordered phase, or an upward curvature
characteristic of the  ordered phase,
depending on whether the temperature is higher or lower than
a critical value. In 4D, the chiral-glass transition temperature
estimated in this way is
$T_{\rm CG}/J=0.38(2)$, while in 5D it is $T_{\rm CG}/J=0.60(2)$. Both are
close to our estimate above based on the chiral Binder ratio.
Our observation that $C_{\chi}(t)$ exhibits an upward curvature below
$T_{\rm CG}$ indicates that the chiral-glass ordered state
is ``rigid'' with a nonzero long-range order parameter.

 From the above analysis,
the occurrence of the chiral-glass LRO in 4D and 5D seems now well
established.
The next question is
whether the chiral-glass
order accompanies the standard SG order. If the SG order occurs at the
same temperature as the chiral-glass order, the transition is likely to
be of the standard type, at least in the sense that the order parameter of
the transition is the spin, not the chirality. (Here, recall that the
SG ordered state in the Heisenberg SG inevitably accompanies the chiral-glass
order as long as the spin is frozen in a noncoplanar manner.)
By contrast, if the SG order occurs at a temperature below
the ordering temperature of the
chirality, it means the unusual situation, {\it i.e.\/},
the occurrence of the spin-chirality decoupling,
the pure chiral-glass transition and the pure chiral-glass ordered state.
To clarify this issue, we examine the spin Binder ratio in the next subsection.

\subsection{Spin Binder ratio}
\begin{figure}[ht]
\leavevmode
\begin{center}
\begin{tabular}{l}
\psfrag{arrow}{{\Large $\uparrow$}}
\psfrag{arrow2}{{\Large $\uparrow$}}
\includegraphics[scale=0.7]{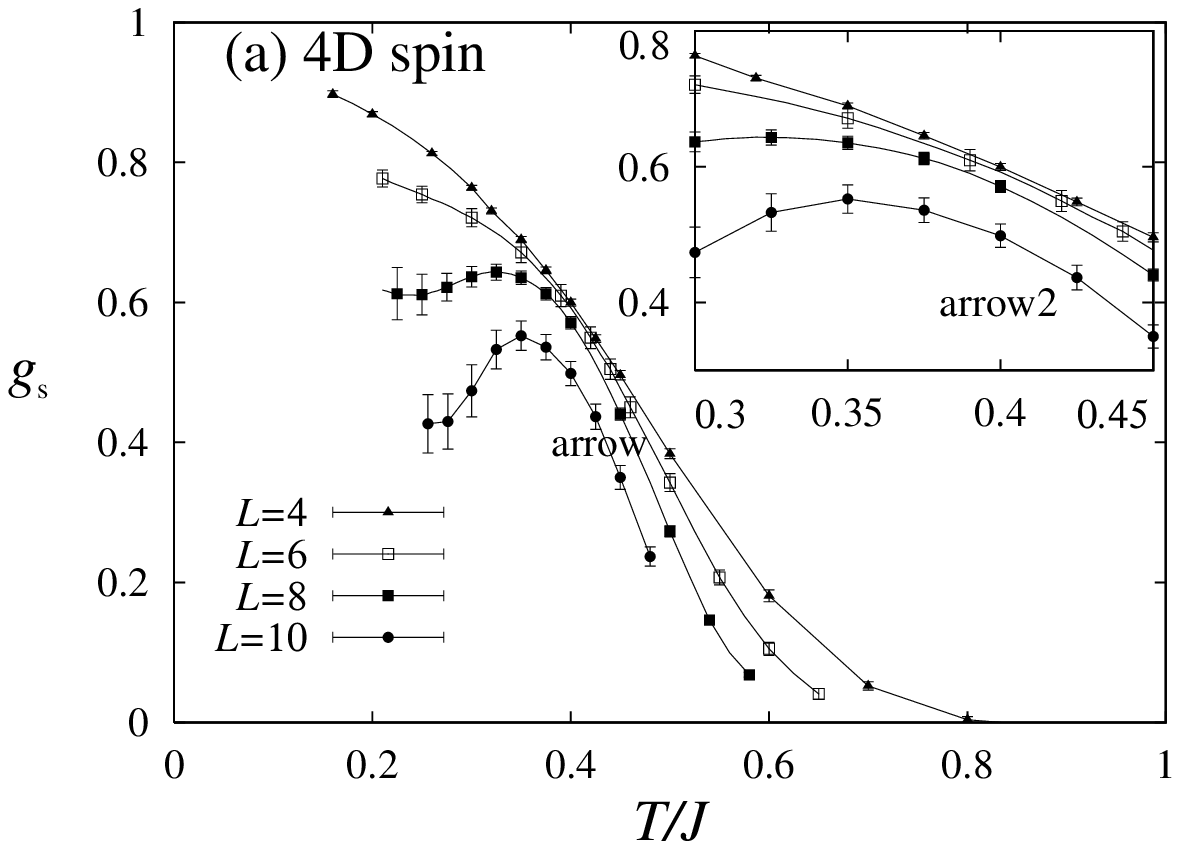}
\end{tabular}
\begin{tabular}{ll}
\psfrag{arrow}{{\Large $\uparrow$}}
\psfrag{arrow2}{{\Large $\uparrow$}}
\includegraphics[scale=0.7]{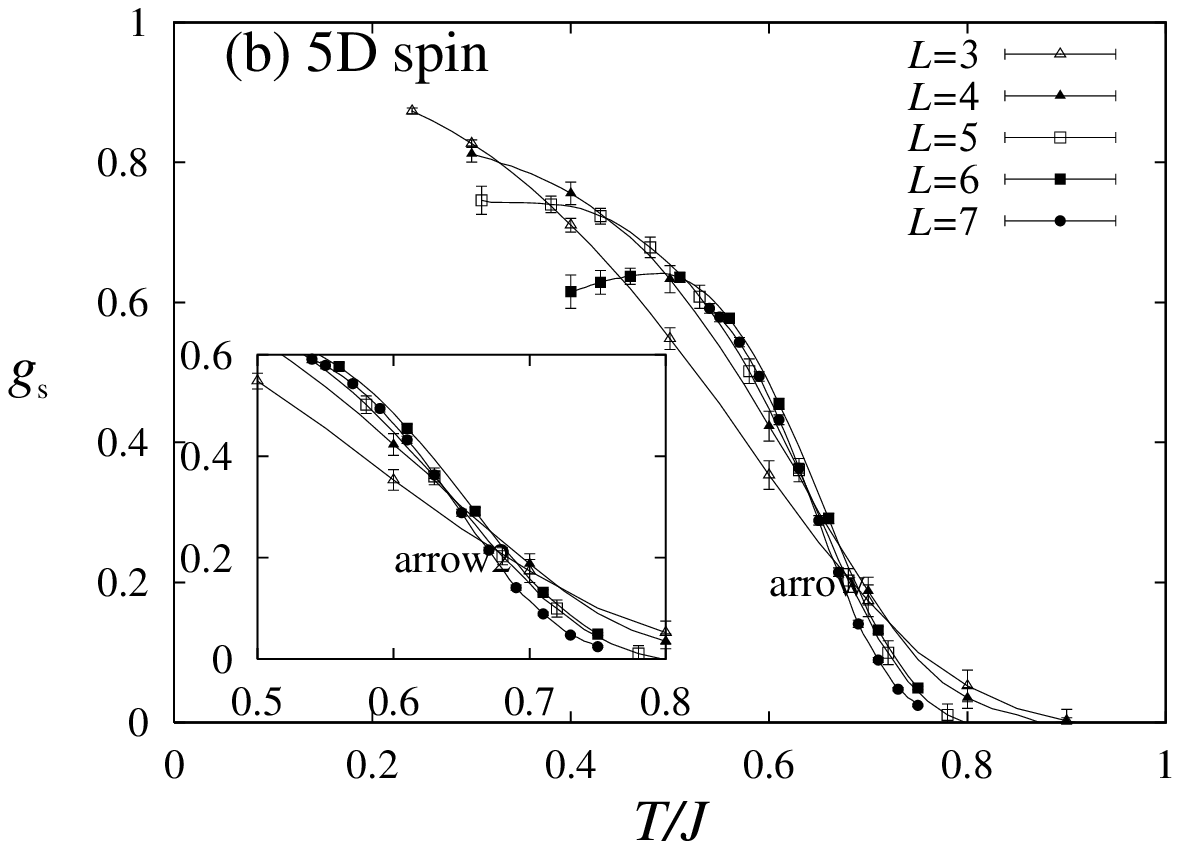}
&
\psfrag{arrow}{{\Large $\downarrow$}}
\psfrag{arrow2}{{\Large $\downarrow$}}
\includegraphics[scale=0.7]{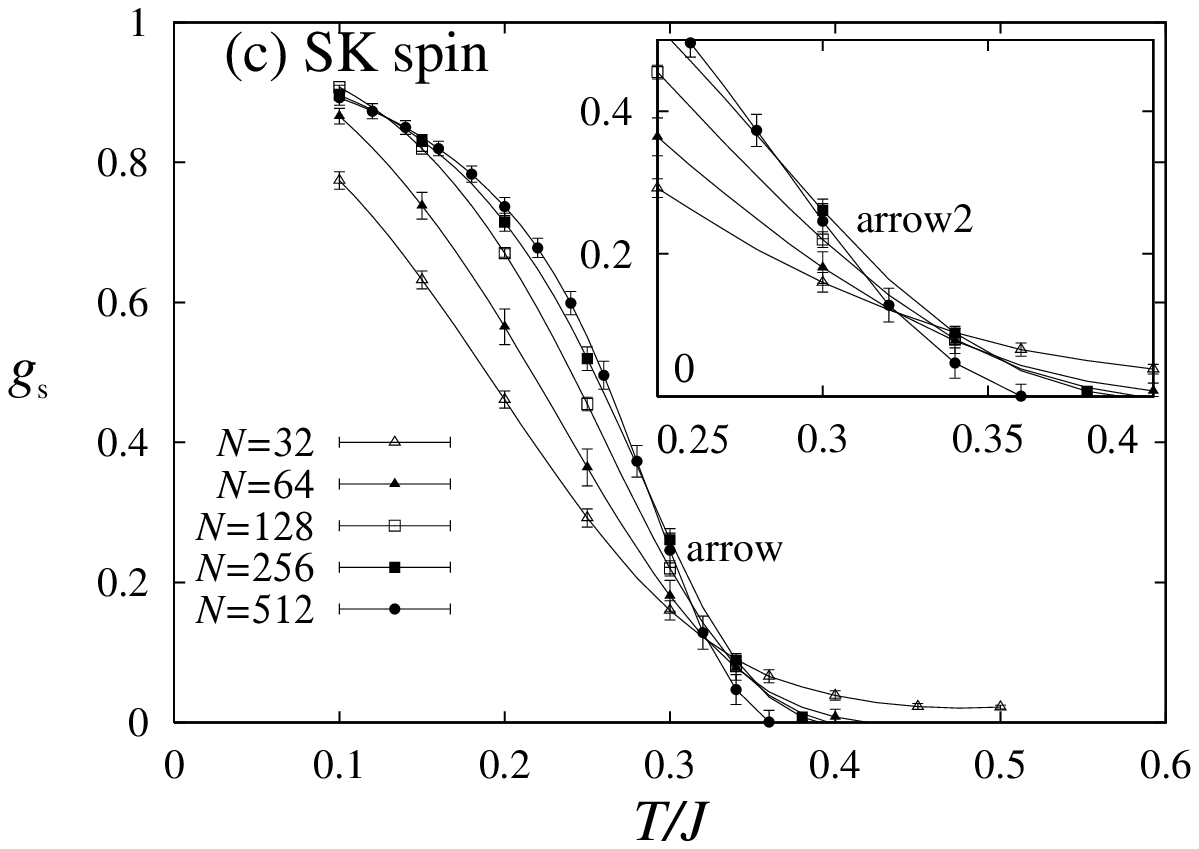}
\end{tabular}
\caption{Temperature and size dependence of the spin Binder ratio $g_{\rm s}$
in 4D (a), in 5D (b), and in the SK model (c).
Magnified figures are given as insets. The arrow in each figure represents
the location of the chiral-glass transition point.
}
\label{fig_gs}
\end{center}
\end{figure}

In \figtag\ref{fig_gs}, we show the temperature and size
dependence of the
Binder ratio of the spin in the cases of
(a) 4D, (b) 5D and (c) the SK model.
In each figure, magnified figure is embedded to show the
detailed behavior of $g_{\rm s}$ in the region of interest.
The arrow in each figure indicates the location of the transition
point of the chirality determined above.
In 4D,
as can be seen from \figtag\ref{fig_gs} (a), $g_{\rm s}$ for the range of sizes
$4\leq L\leq 8$ appears to almost merge
at a temperature $T/J\simeq 0.4$ close to the chiral-glass transition
temperature determined above.
This seems to suggest that the spin sector
also becomes critical at $T\simeq T_{{\rm CG}}$, which may
indicate the simultaneous spin-glass and chiral-glass transition at
$T/J\simeq 0.4$. Quite remarkably, however,
the spin Binder ratio $g_{\rm s}$ for our
largest size $L=10$ comes definitely below the curves for $L\leq 8$. In fact,
the $L=10$ curve lies below the $L\leq 8$ curves more than four sigmas, and
the observed departure is statistically well significant.
 From this observation, we conclude that, although the spin sector once becomes
near-critical at the chiral-glass transition $T=T_{{\rm CG}}$
on short length scales, it eventually remains off-critical (paramagnetic) at
$T=T_{{\rm CG}}$ on longer length scales. Namely,
the spin-chirality decoupling previously observed in 3D in \reftag~\cite{HK1}
seems to come into play in 4D as well, which keeps the spin sector being
paramagnetic even below $T=T_{{\rm CG}}$.
If this is the case, the transition
in 4D is a pure chiral-glass transition and the ordered state is a
pure chiral-glass state not accompanying the standard SG order.
Even in certain temperature range below $T_{{\rm CG}}$,
$g_{\rm s}$ is expected to approach zero
in the $L\rightarrow \infty$ limit,
which seems consistent with the present data. We interpret the
strange structure observed in
$g_{\rm s}$ near $T=T_{{\rm CG}}$ for larger $L$ as a remanence of the near-critical
behavior of the spin at the chiral-glass transition.
Meanwhile, due to the lack of our data in the lower temperature region
$T/J\lsim 0.2$,
it is difficult to determine from
the present data whether the SG transition occurs either
only at zero temperature, $T_{{\rm SG}}=0$, or at a finite temperature
below the chiral-glass transition
temperature, $0<T_{{\rm SG}}<T_{{\rm CG}}$. Nevertheless, \figtag\ref{fig_gs}(a)
strongly suggests that $T_{{\rm SG}}$, if it is nonzero, is less than $0.2J$.

We note in passing that the spin Binder ratio $g_{\rm s}$ of the same 4D
model was calculated by Coluzzi for smaller sizes
$L=3,4$ and $5$ and at higher temperatures $T/J\geq 0.5$~\cite{Coluzzi}.
She observed that $g_{\rm s}$ for $L=3,4$ and $5$ appeared
to merge
around $T/J\simeq 0.5$, and suggested that there occurred
a standard SG transition at $T/J\simeq 0.5$. In the present calculation
made for larger lattice sizes and for lower temperatures,
although we indeed observed
a near-merging behavior of $g_{\rm s}$, it occurred at a
temperature $T/J\simeq 0.4$ near $T_{{\rm CG}}$,
somewhat lower than the estimate of \reftag~\cite{Coluzzi},
and most importantly, $g_{\rm s}$ for larger lattices show clear deviation
from the merging behavior, suggesting that the Heisenberg spin remains
paramagnetic even below the chiral-glass transition point.

Now, we turn to the spin Binder ratio in 5D.
As shown in \figtag\ref{fig_gs} (b),
$g_{\rm s}$ for $3\leq L\leq 7$ show a crossing at a
temperature around $T/J\simeq 0.60$,
strongly suggesting that the SG order occurs at $T_{{\rm SG}}/J=0.60(2)$.
At lower temperatures, $g_{\rm s}$ for larger $L$ tends to come down again,
exhibiting a behavior reminiscent to the one observed in 4D.
In particular, at low enough temperatures, $g_{\rm s}$
decreases with $L$. The observed non-monotonic
temperature dependence and the peculiar size dependence of
$g_{\rm s}$ below $T_{{\rm SG}}$ suggests that the SG state below
$T_{{\rm SG}}$ might be a nontrivial one
accompanied by a peculiar RSB, possibly the one-step-like one as observed
in the chirality sector.

In the SK case, the calculated $g_{\rm s}$ exhibits a clear crossing
behavior at around $T/J=1/3$, an exactly known SG transition temperature of the
model. At lower temperatures,
$g_{\rm s}$
monotonically increases with $N$ in contrast to the 4D and 5D cases,
and eventually appear to converge to the temperature-dependent
values less than unity.
The observation that the asymptotic
$g_{\rm s}(L\rightarrow \infty)$ in the region $0<T<\frac{1}{3}J=T_{\rm SG}$
differs from unity reflects the fact that the SG ordered state of
the SK model accompanies
a full (or hierarchical) RSB.

\subsection{Chiral overlap distribution function}

\begin{figure}[ht]
\leavevmode
\begin{center}
\begin{tabular}{l}
\includegraphics[scale=0.7]{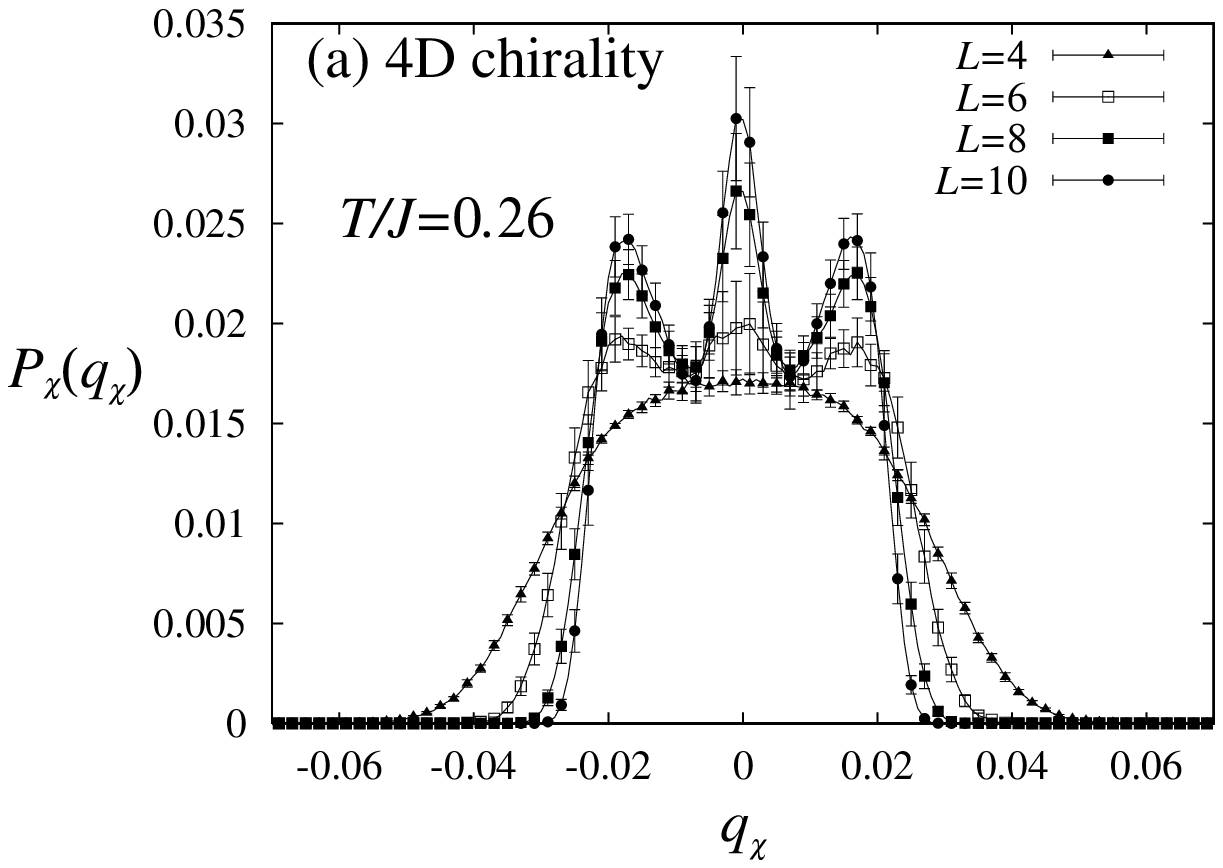}
\end{tabular}
\begin{tabular}{ll}
\includegraphics[scale=0.7]{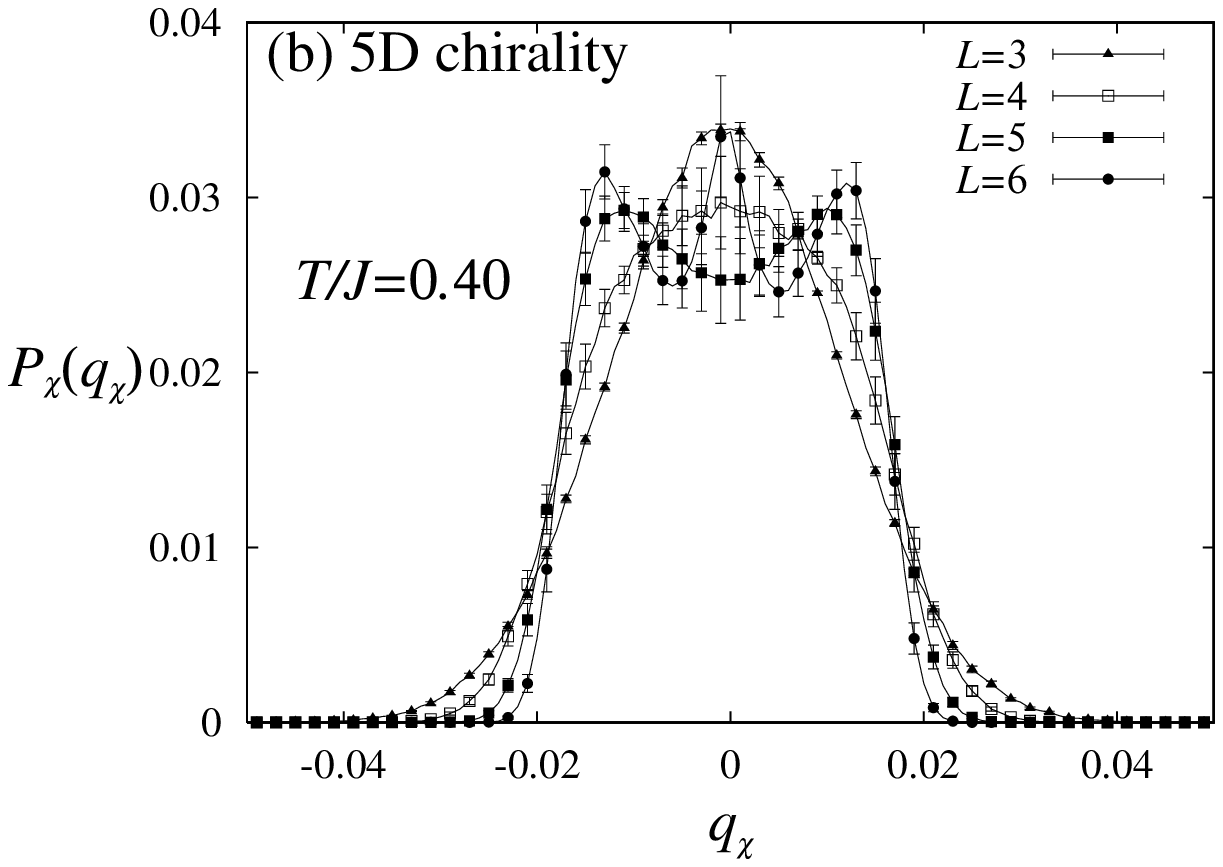}
&
\includegraphics[scale=0.7]{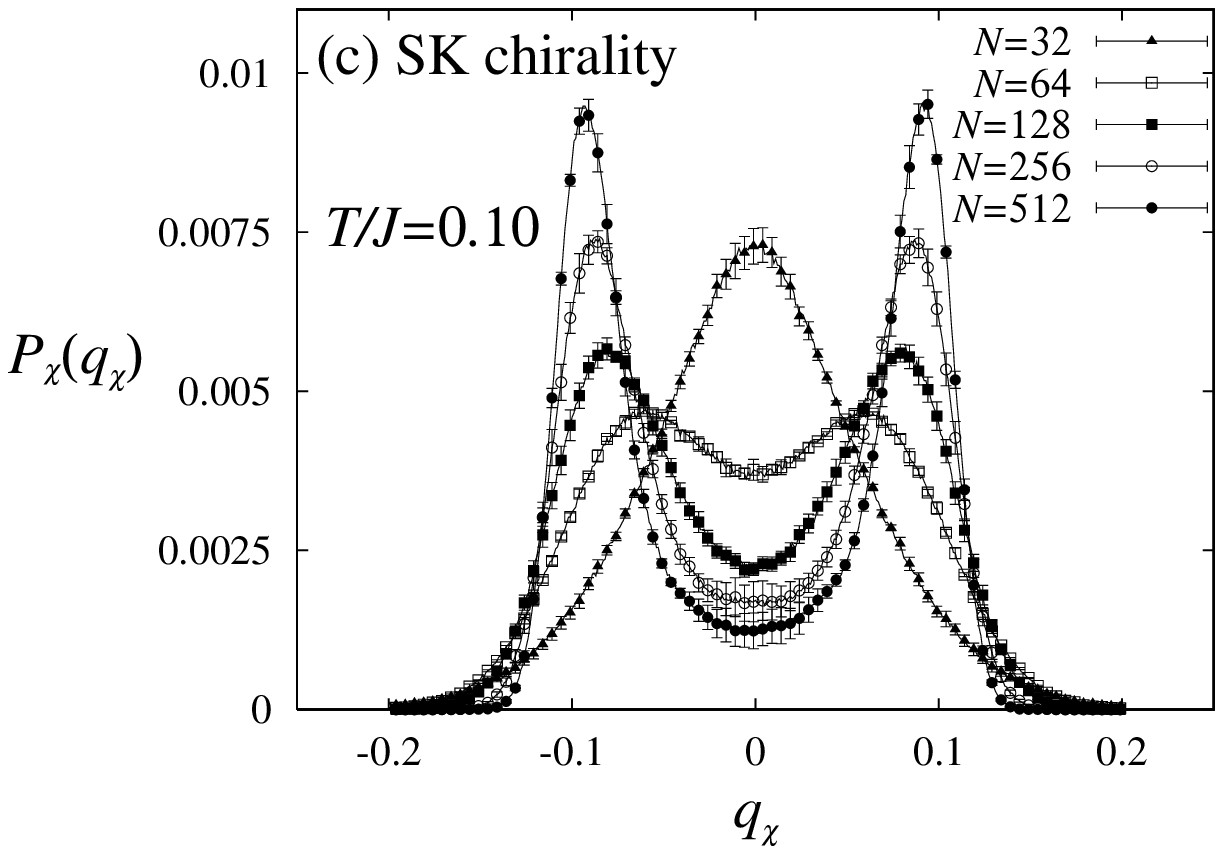}
\end{tabular}
\caption{Overlap distribution function of the chirality,
in 4D at $T/J=0.26$ (a), in 5D at $T/J=0.40$ (b), and in the
SK model at $T/J=0.10$ (c).}
\label{fig_Pqx}
\end{center}
\end{figure}

In \figtag\ref{fig_Pqx},
we show the overlap distribution function of the
chirality, $P_{\chi}(q_{\chi})$,
well below $T_{\rm CG}$, at around
$T\simeq \frac{2}{3}T_{{\rm CG}}$.
In 4D, in addition to the "side peaks" corresponding to
$q_{\chi}=\pm q_\chi^{\rm EA}$ which grow and  sharpen with
increasing $L$,
a "central peak" appears at $q_{\chi}=0$ for $L\geq 6$, which also
sharpens and gets higher with increasing $L$.
These features
were reminiscent to the ones
observed in 3D, which was
interpreted as a signature of the
one-step-like RSB in the chiral-glass ordered
state~\cite{HK1,3DHSGinH_1,3DHSGinH_2}.
Our present data suggest that the chiral-glass state in 4D also accompanies
the one-step-like RSB as in the 3D case. This is fully consistent with the
observed behavior of the chiral Binder ratio.

A central peak in $P_{\chi}(q_{\chi})$ is also observed
in 5D in the largest lattice size $L=6$. This  suggests the
occurrence of the one-step-like RSB also in 5D.

In the SK case, the calculated
$P_{\chi}(q_{\chi})$ exhibits
the side peaks at $q_{\chi}=\pm q_\chi^{\rm EA}$ only, without
a central peak at $q_{\chi}=0$ for any size studied. Instead,
the value of $P_{\chi}(0)$ gradually decreases
with increasing $N$, where $1/N$-extrapolation of $P_{\chi}(0)$
to $N\rightarrow\infty$ gives
a nonzero value,
$P_{\chi}(q_{\chi}=0, N=\infty)\simeq 6.0\times 10^{-4}$.
Thus, in the SK case, the
chirality exhibits the standard full RSB, in apparent
contrast to the 4D and 5D cases.

\subsection{Spin overlap distribution function}
\begin{figure}[ht]
\leavevmode
\begin{center}
\begin{tabular}{l}
\includegraphics[scale=0.7]{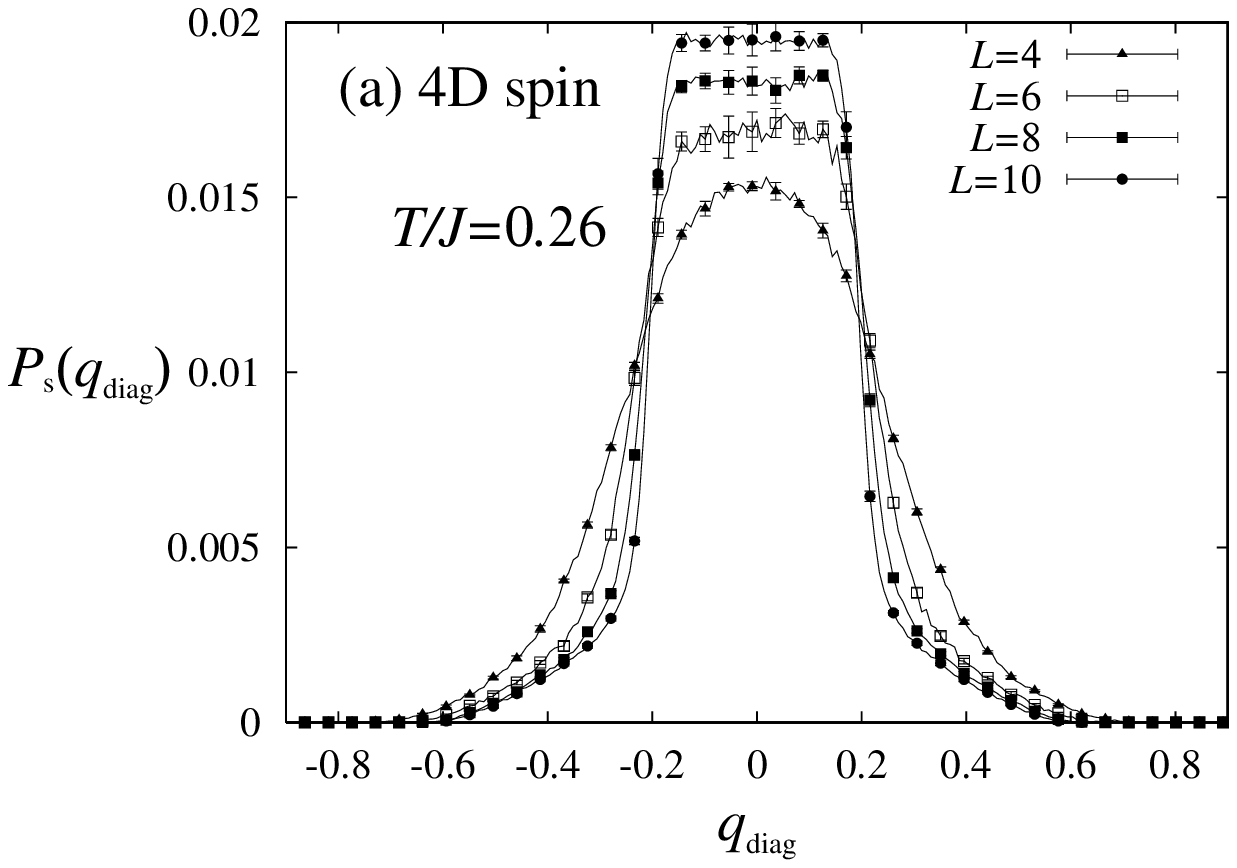}
\end{tabular}
\begin{tabular}{ll}
\includegraphics[scale=0.7]{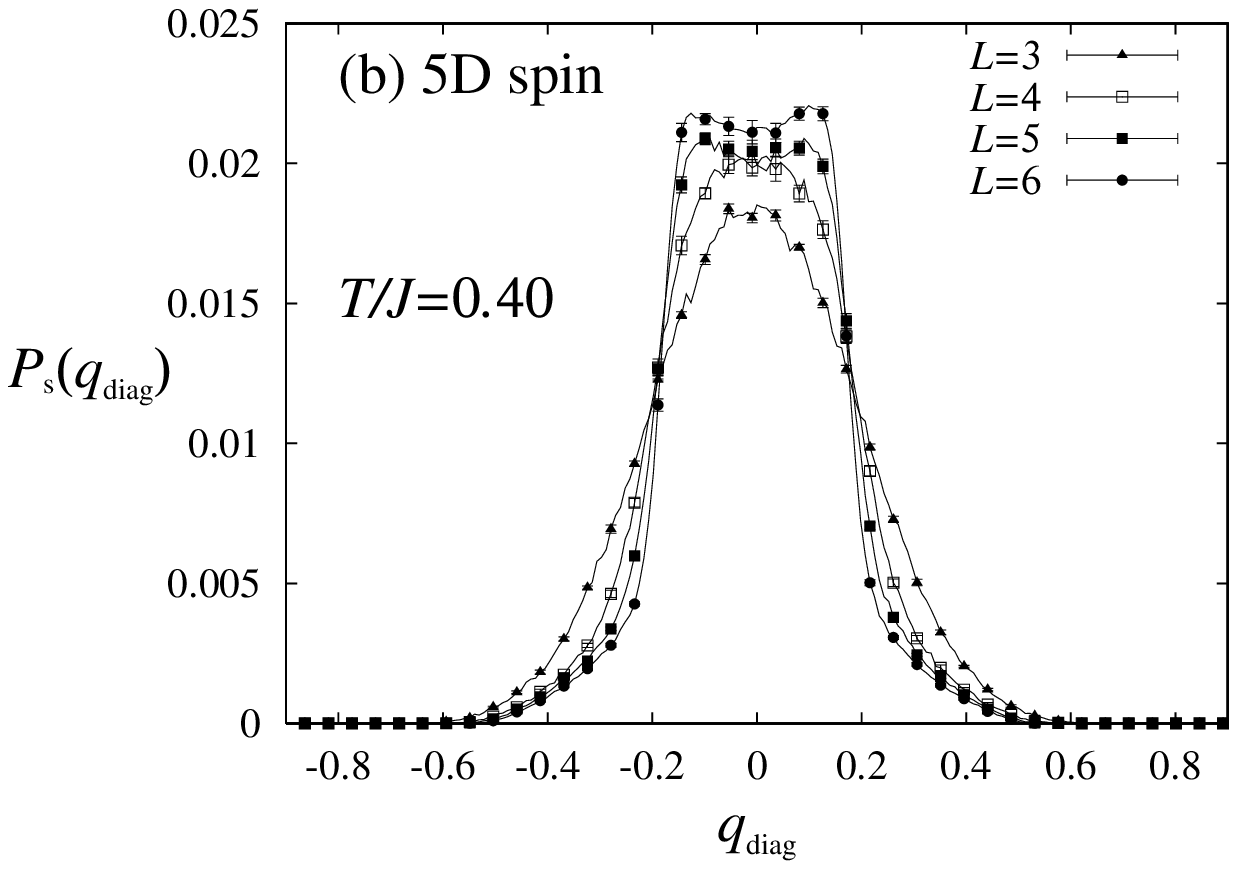}
&
\includegraphics[scale=0.7]{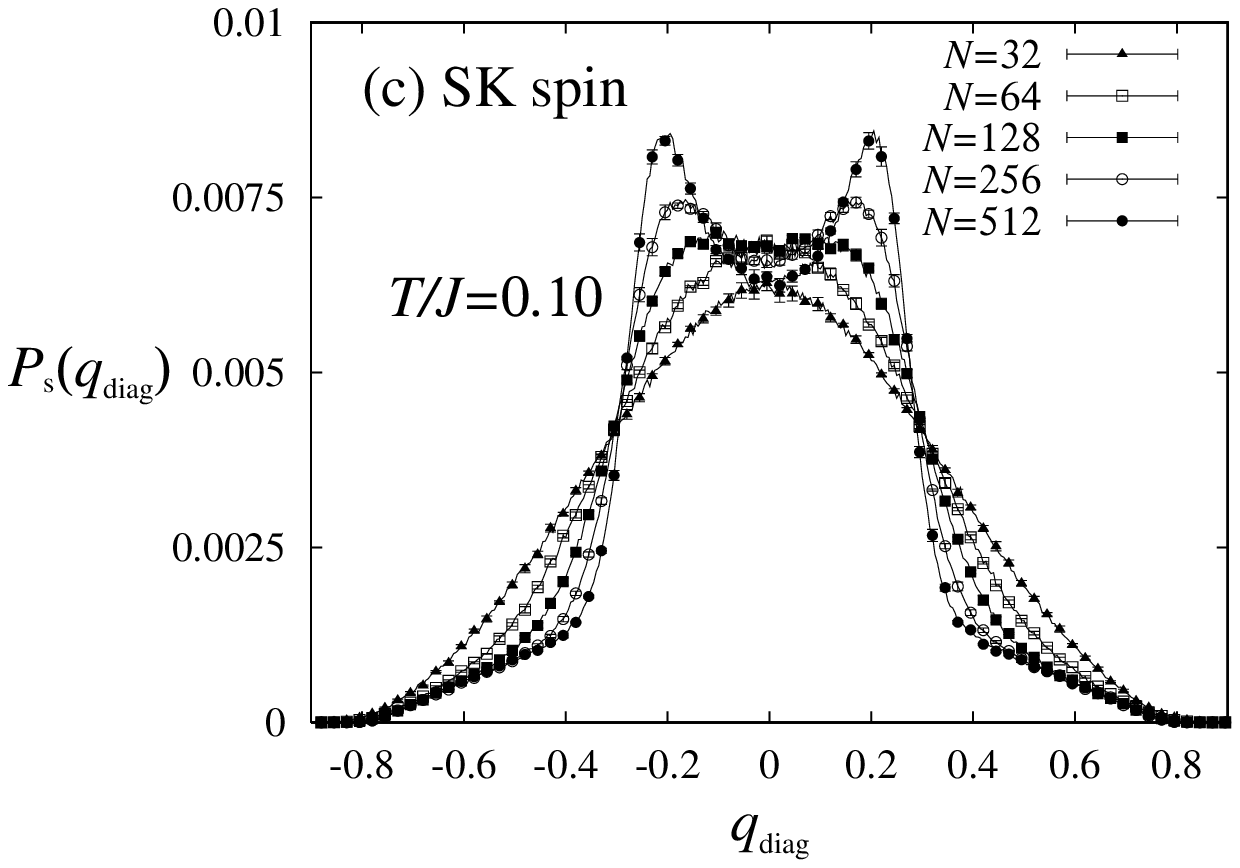}
\end{tabular}
\caption{Overlap distribution function of the diagonal part of
the spin, in  4D at $T/J=0.26$ (a), in 5D at $T/J=0.40$ (b),
and in the SK model at $T/J=0.10$ (c).}
\label{fig_Psqd}
\end{center}
\end{figure}

In \figtag\ref{fig_Psqd}, we show the diagonal spin-overlap
distribution function, $P_{\rm s}(q_{\rm diag})$, for the cases of
(a) 4D, (b) 5D and (c) the SK model. The temperatures are taken to be
the same as those for the chiral overlap distribution shown
in \figtag\ref{fig_Psqd},
{\it i.e.\/}, about $\frac{2}{3}T_{{\rm CG}}$.

In the SK case shown in \figtag\ref{fig_Psqd}(c),
the shape of $P_{\rm s}(q_{\rm diag})$ is
similar to the one of \figtag\ref{fig_LRO}, with symmetric diverging peaks
observed at $q_{\rm diag}\simeq \pm 0.2$. These peaks are then identified with
the ones expected at $\pm \frac{1}{3}q_{\rm s}^{\rm EA}$ when there is
a finite SG LRO. This observation is fully consistent with
the fact that the standard SG LRO with a nonzero
$q_{\rm s}^{\rm EA}$ is realized in the ordered state of the SK model.
In 5D, the growing symmetric peaks also appear
at $q_{\rm diag}\simeq \pm 0.18$ for $L\geq 5$,
suggesting that the SG LRO characterized by
a nonzero $q_{\rm s}^{\rm EA}$ is realized.

In 4D, by contrast,
no peaks corresponding
to $\pm \frac{1}{3}q_{\rm s}^{\rm EA}$ are observed, at least
within the range of sizes we simulate.
Instead, $P_{\rm s}(q_{\rm diag})$ exhibits a marginal behavior,
staying nearly flat with a plateau-like structure at
$|q_{\rm diag}|\lsim 0.2$
for $L\geq 6$. With increasing $L$,
this plateau gradually gets higher but no side peaks show up.
Since the normalization condition of $P_{\rm s}(q_{\rm diag})$ inhibits
the plateau of finite width getting higher indefinitely, one plausible
asymptotic behavior of $P_{\rm s}(q_{\rm diag})$ might be that it
eventually converges in the $L\rightarrow \infty$ limit
to the Gaussian distribution around $q_{\rm diag}=0$.
In that sense,
the observed behavior is consistent with the spin disorder at
this temperature.
However, solely from the present data,
we cannot completely rule out
the possibility that the $\pm q_{\rm s}^{\rm EA}/3$ peaks
characteristic of the SG LRO eventually show up for still larger $L$.
Anyway, in the range of sizes studied $L\leq 10$,
we have observed no sign of such side
peaks, in contrast to the 5D case where the side peaks appear already
for $L=5$. Another possibility may be that the 4D model exhibits a
finite-temperature SG transition but the SG state is a critical phase
with a vanishing SG order parameter $q_{\rm s}^{\rm EA}=0$, as
expected for the system at its LCD. However, as argued below, such a
LCD behavior is not supported from our data of the critical properties.

\subsection{Chiral-glass susceptibility}
\begin{figure}[ht]
\leavevmode
\begin{center}
\begin{tabular}{l}
\includegraphics[scale=0.7]{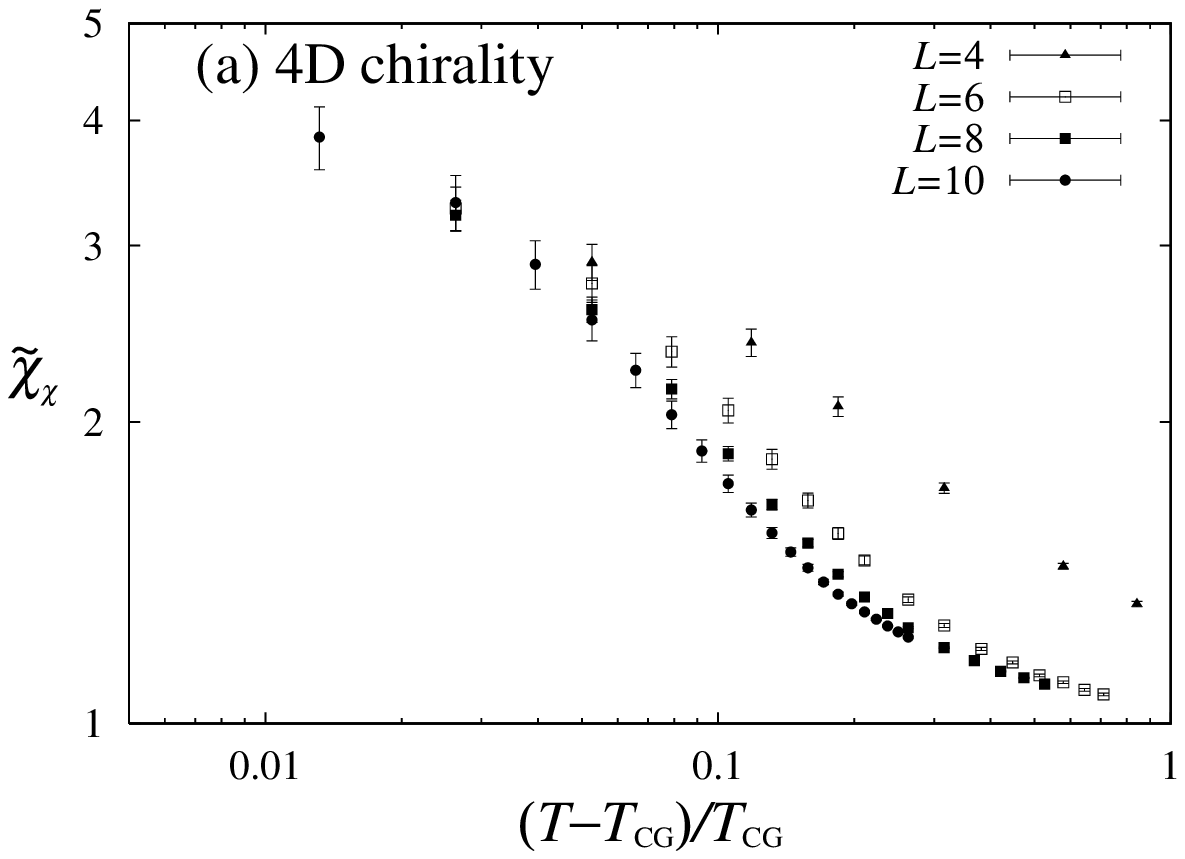}
\end{tabular}
\begin{tabular}{ll}
\includegraphics[scale=0.7]{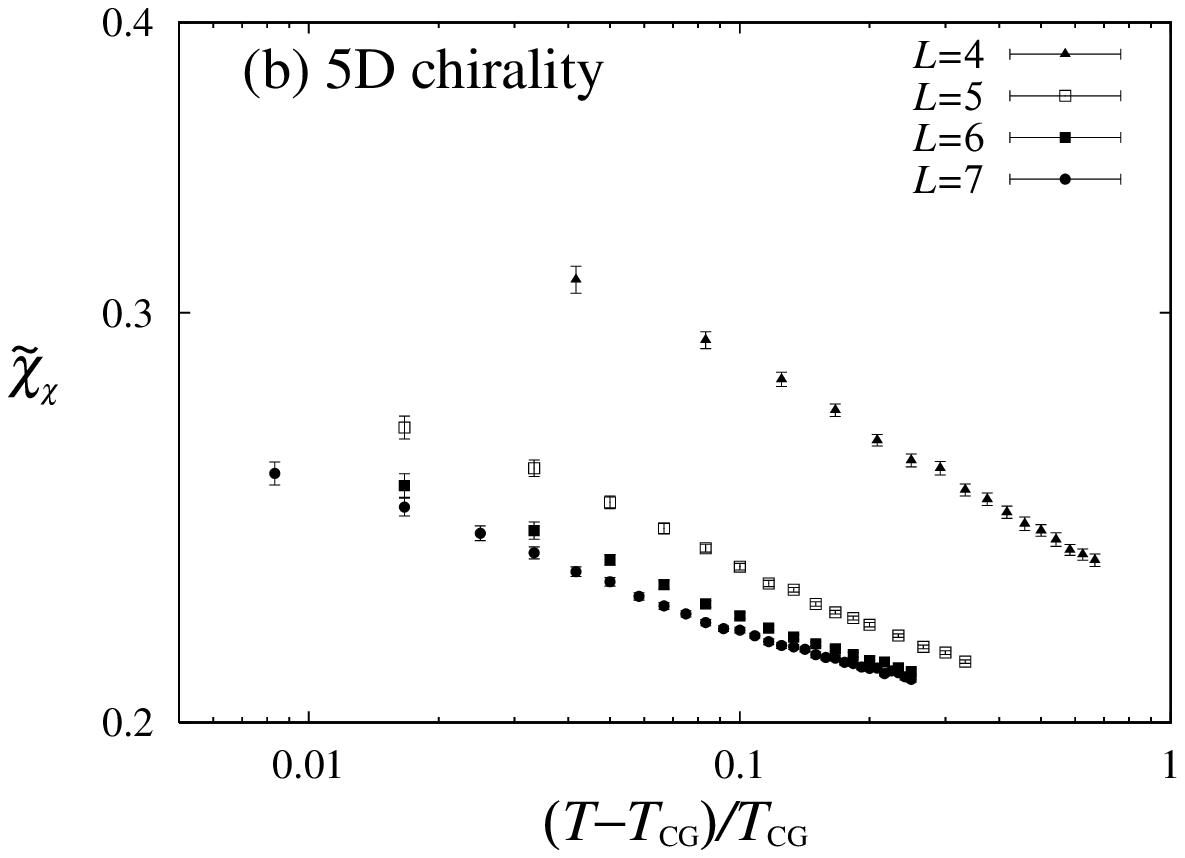}
&
\includegraphics[scale=0.7]{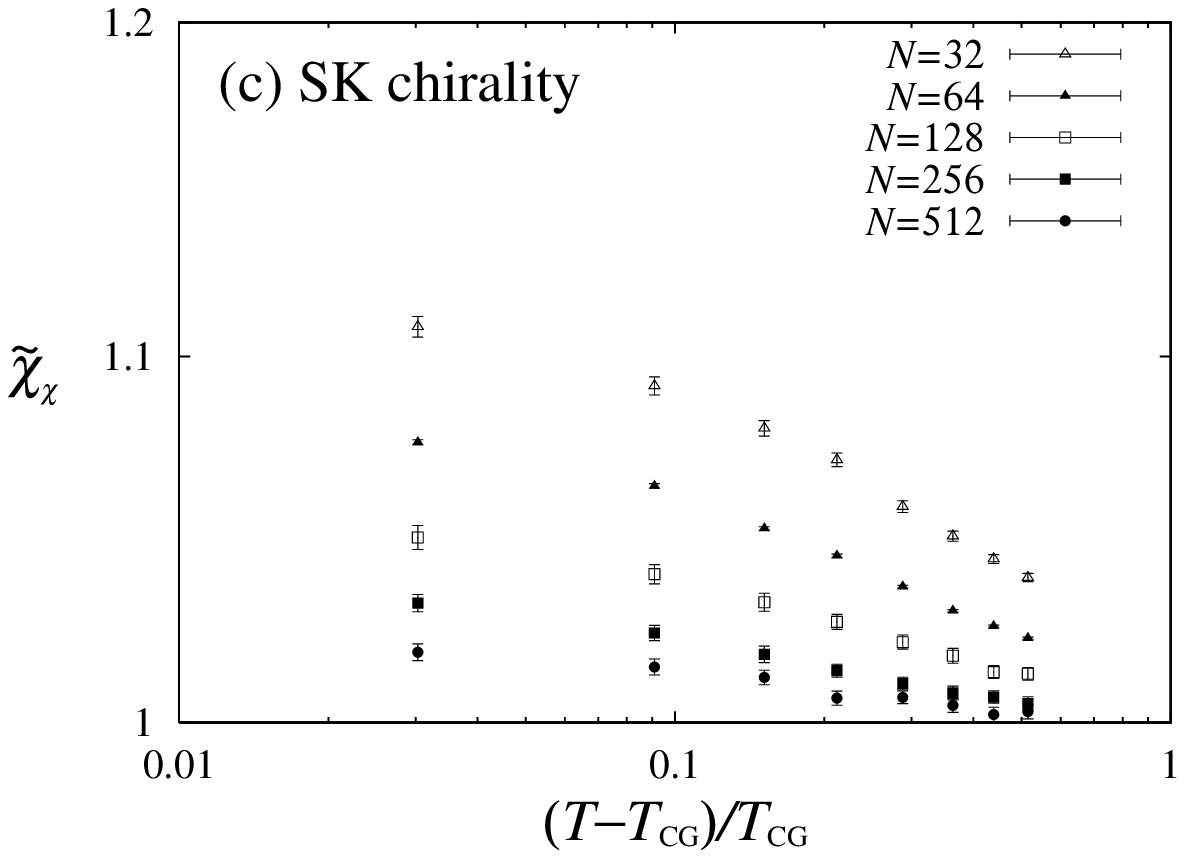}
\end{tabular}
\caption{Temperature and size dependence of the reduced chiral-glass
susceptibility in 4D (a), in 5D (b), and in the SK model (c).}
\label{fig_Xcg}
\end{center}
\end{figure}

%
 From the data presented above, we have concluded
that in 4D the chirality is an order parameter of the transition,
but not so in 5D and in the SK model: In the latter cases,
the order parameter of the transition is the spin, while the chirality
order is parasitic to the spin order. In order to
examine further the validity
of such a picture, we show in \figstag\ref{fig_Xcg} the
temperature and size dependence of the
reduced chiral-glass susceptibility, where $\tilde{\chi}_\chi$ is plotted
as a function of the reduced temperature $(T-T_{{\rm CG}})
/T_{{\rm CG}}$  on a log-log scale.
As determined above,
$T_{{\rm CG}}$ is taken to be $T_{{\rm CG}}/J=0.38$ in 4D
and $T_{{\rm CG}}/J=0.60$ in 5D.
In the SK case, we put $T_{{\rm CG}}/J=\frac{1}{3}$  which is exact.

As can be seen from \figstag\ref{fig_Xcg} (b) and (c),
the reduced chiral-glass susceptibilities of the 5D and the SK models
do not exhibit any sign of divergence: They stay small at any temperature
$T>T_{{\rm CG}}$, and most notably, $\tilde{\chi}_\chi$ gets smaller with increasing
the system size $L$. Such a size dependence is completely opposite to the
one expected for a diverging quantity in the critical region.
We note that
even at temperatures close to the transition temperature, no
sign of the reversal of the size dependence
is discernible. In fact, in the SK case,
such a non-diverging behavior of $\tilde{\chi}_\chi$
is just as expected. In the SK model, the standard SG exponents are known
to be $\alpha=-1$, $\beta_{{\rm SG}}=1$
and $\gamma_{{\rm SG}}=1$. Then the chiral-glass
exponent $\beta_{{\rm CG}}$ is expected to be $\beta_{{\rm CG}}=3$, because
the chirality is cubic in the spin variables. Then, the chiral-glass
susceptibility exponent is obtained as $\gamma_{{\rm CG}}=-3$ from
the scaling relation $\alpha+2\beta_{{\rm CG}}+\gamma_{{\rm CG}} =2$.
Negative $\gamma_{{\rm CG}}$ means that the chiral-glass
susceptibility of the SK model does not diverge at the SG transition.
Very much similar behavior observed in 5D suggests that the
the chiral-glass susceptibility of the 5D model
does not diverge at the transition, either.
Hence, our observations for $\tilde{\chi}_\chi$
are fully consistent
with our previous finding that the order parameter of the transition
in 5D and in the SK model is the
spin, not the chirality.

By contrast, in 4D, $\tilde{\chi}_\chi$ exhibits a different behavior.
Although in the investigated temperature regime
$\tilde{\chi}_\chi$ stays rather small and tends
to decrease with increasing $L$,
similarly to the behavior observed in 5D and in the SK limit,
its size dependence is about to change
in a close vicinity of $T_{{\rm CG}}$. More specifically, the $L=10$ data
catch up the $L=8$ data at $t\simeq 0.04$, and at temperatures further
close to $T_{{\rm CG}}$, exceeds the $L=8$ data, where the data show
significant finite-size rounding preventing the observation of
the asymptotic critical behavior. This suggests that the
critical region of the chiral-glass transition might be very narrow in 4D,
limited to the regime $t\lsim 10^{-2}$. In the temperature range
outside this,  the chirality
exhibits a mean-field-like non-diverging behavior similar to the
one of the SK model. Although
we cannot directly get into this narrow critical region in our
present simulation due to the computational limitation,
the observed behavior of $\tilde{\chi}_\chi$ of the 4D model
indeed suggests that such a crossover from the mean-field-like behavior
at $t\gsim 10^{-2}$ to the diverging critical behavior
at $t\lsim 10^{-2}$ does occur. Unfortunately,
inaccessibility to the asymptotic critical region prevents us from
estimating the chirality exponents. In order to estimate
the asymptotic chiral critical exponents, one needs to approach
the temperature regime $t\lsim 10^{-2}$ with larger lattices $L\geq 10$,
which is not feasible with the computational capability presently available
to us.

\subsection{Critical Exponents}
\label{subsecCritical}
In this subsection, we analyze the critical properties of the
SG transition in each case of 5D and of the SK model.
In the case of 4D, our analysis has suggested that
the transition is a pure chiral-glass transition,
the associated chiral critical regime being
narrow, $t\lsim 10^{-2}$, which prevents us from
estimating  the chiral critical exponents.

\begin{figure}[ht]
\begin{center}
\includegraphics[scale=0.7]{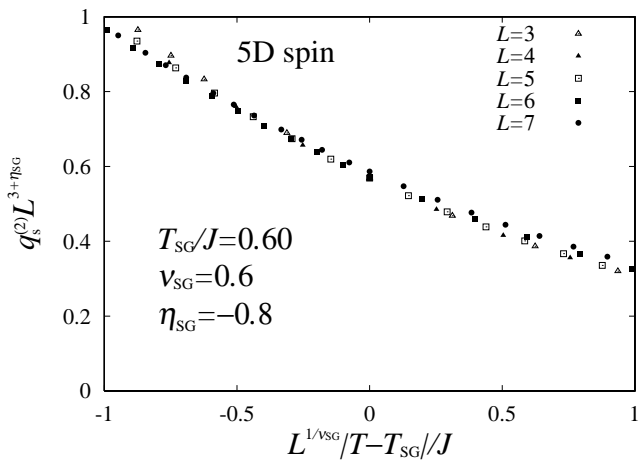}
\caption{Finite-size scaling plot of the spin-glass order parameter in 5D.}
\label{fig_FSS}
\end{center}
\end{figure}

In 5D, we have concluded that
the transition is the standard SG transition, {\it i.e.\/},
the order parameter is the spin, not the chirality, although
the chirality also takes a nonzero value in the SG ordered state
reflecting the noncoplanar character of the spin order.
We estimate the associated SG exponents via the standard
finite-size-scaling analysis of $q_{{\rm s}}^{(2)}$, based on the relation,
\begin{equation}
q_{\rm s}^{(2)}\approx L^{-(d-2+\eta_{\rm SG})}
f(L^{1/\nu_{\rm SG}}|T-T_{{\rm SG}}|)\ \ ,
\end{equation}
where the $T_{\rm SG}$ value is set to the best value
determined above, $T_{\rm SG}/J=0.60$.
The best estimates of the SG exponents are
$\nu_{\rm SG}=0.6(2)$ and $\eta_{\rm SG}=-0.8(2)$: See \figtag\ref{fig_FSS}.
 From the standard scaling relation, we get other exponents as
$\alpha=-1.0(5), \beta_{{\rm SG}}=0.7(3)$ and $\gamma_{{\rm SG}}=1.7(5)$.
One sees that these exponents are not far from the mean-field
exponents expected above the upper critical dimension $D=6$.

\begin{figure}[ht]
\begin{center}
\includegraphics[scale=0.7]{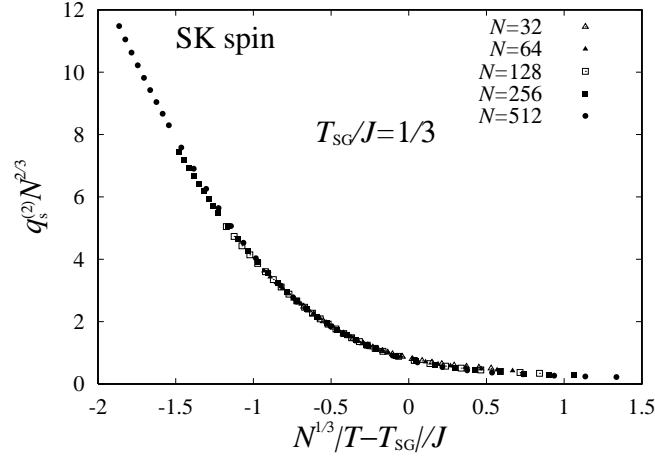}
\caption{Finite-size scaling plot of the spin-glass order parameter in
the SK model.}
\label{fig_FSS_SK}
\end{center}
\end{figure}

In the SK case, mean-field exponents should be exact. Indeed,
as shown in \figtag\ref{fig_FSS_SK}, our data of $q_{{\rm s}}^{(2)}$ are
entirely consistent with such a mean-field behavior, $\beta_{{\rm SG}}=1$ and
$\eta_{{\rm SG}}=0$~\cite{InfSys_FSS,KH_RSB}.

\begin{figure}[ht]
\leavevmode
\begin{center}
\begin{tabular}{ll}
\includegraphics[scale=0.7]{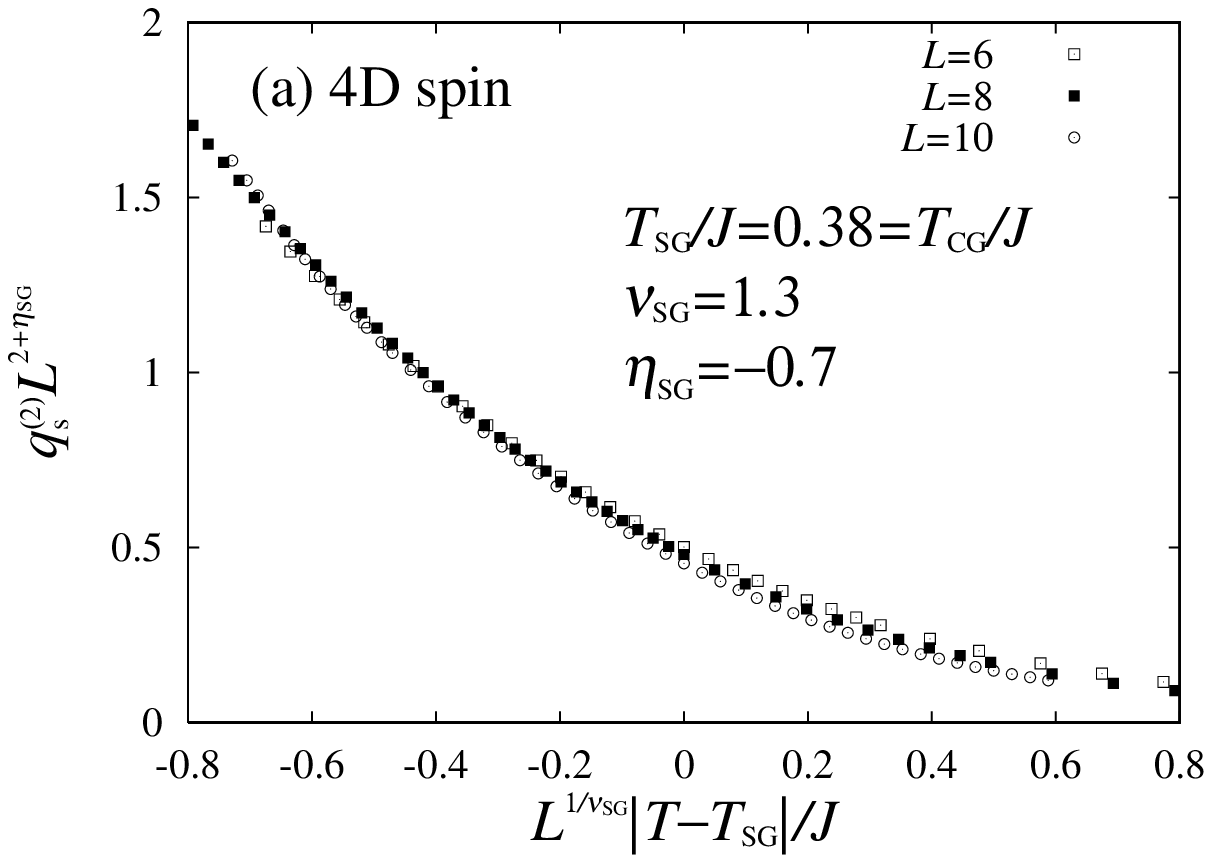}
&
\includegraphics[scale=0.7]{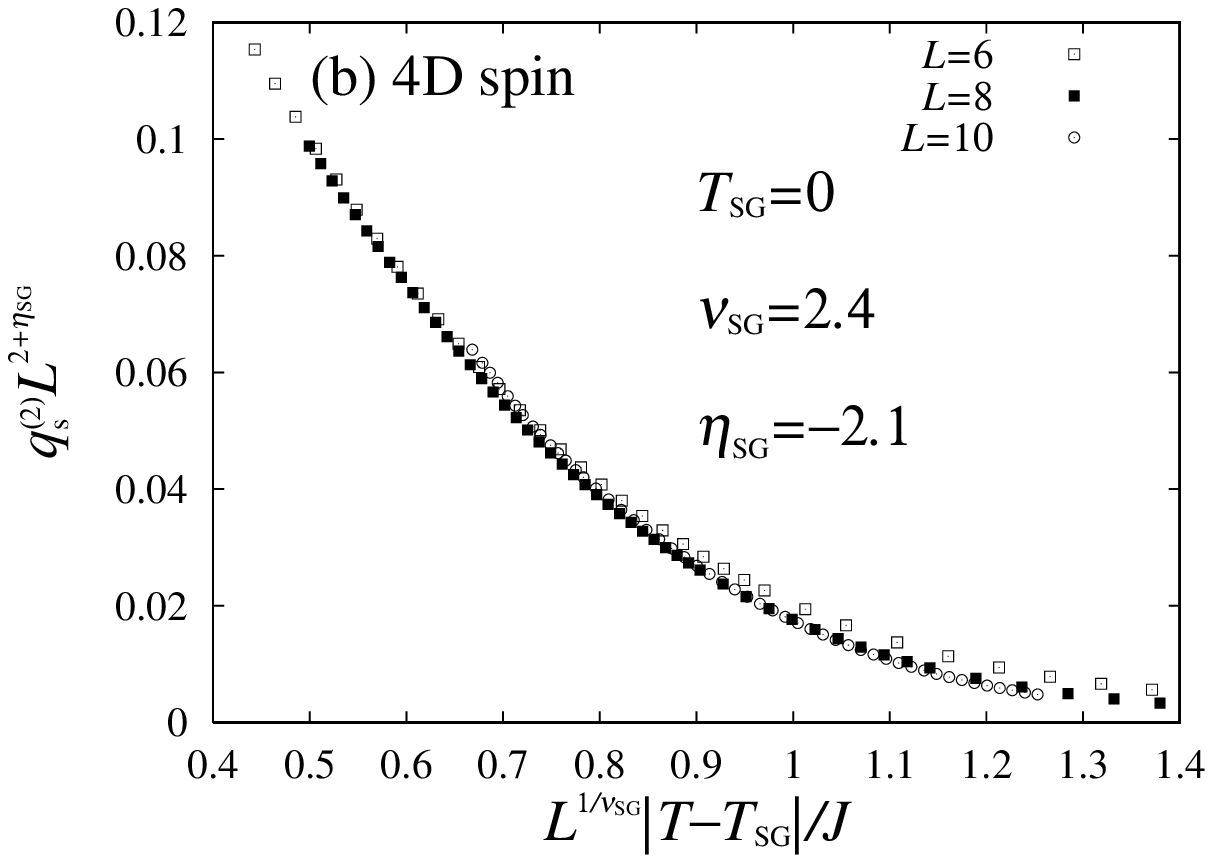}
\end{tabular}
\caption{Finite-size scaling plot of the spin-glass order parameter in 4D,
assuming (a) $T_{\rm SG}/J=T_{\rm CG}/J=0.38$ and (b) $T_{\rm SG}=0$. In
(b), our best value of $\eta_{{\rm SG}}$ is reasonably close to the exact
value of $\eta_{{\rm SG}}=-2$ expected for the $T=0$ transition.}
\label{fig_FSS_4d}
\end{center}
\end{figure}

In concluding this section, we touch upon the near-critical behavior
of the spin observed around the chiral-glass transition point in 4D.
Although our data of the Binder ratios and the overlap distribution
functions given above have strongly suggested
that the Heisenberg spin remains paramagnetic even below
$T_{{\rm CG}}$ in 4D,
the present $q_{{\rm s}}^{(2)}$ data can be scaled reasonably well
with assuming $T_{{\rm SG}}/J=T_{{\rm CG}}/J=0.38(2)$, at least
from the purely numerical viewpoint.
Such a constrained finite-size scaling analysis of $q_{{\rm s}}^{(2)}$
yields the estimates $\nu '_{{\rm SG}}=1.3(2)$ and $\eta '_{{\rm SG}}=-0.7(2)$:
See \figtag\ref{fig_FSS_4d} (a). Note that the value of $\nu'$ is far
from the LCD value, $\nu'=\infty$.
We also note that the same $q_{{\rm s}}^{(2)}$ data can also be fitted with
a comparable quality by assuming a zero-temperature SG transition,
$T_{{\rm SG}}=0$:
See \figtag\ref{fig_FSS_4d} (b).
As mentioned, we believe that the $\nu '_{{\rm SG}}$ and $\eta '_{{\rm SG}}$
values obtained by assuming $T_{\rm SG}=T_{\rm CG}$ do not represent
true asymptotic exponent values, but just represent {\it effective exponents
describing the short-scale near-critical phenomena which is an echo of the
chiral-glass transition\/}.
Indeed, at short scales, the chirality is never independent of the spin by its
definition.
Hence, the behavior of the spin-correlation related quantities
might well reflect
the critical singularity associated with the {\it chirality\/}
up to certain length and time scales.

\section{Summary and discussion}
\label{Summary}
In summary, we performed a large-scale equilibrium
MC simulation of the 4D, 5D and SK
Heisenberg spin glasses.
In 5D, the model
exhibits a single SG transition at a finite temperature,
reminiscent to the one of the corresponding
mean-field model. Below the transition
temperature $T_{{\rm SG}}/J=0.60(2)$, the spin is frozen in a spatially random
noncoplanar configuration. Although the SG order accompanies
a finite chiral-glass LRO reflecting the noncoplanar nature of the
spin order, the order parameter of the transition is the spin, not the
chirality, and the chiral-glass susceptibility remains nondiverging
at $T=T_{{\rm SG}}$. Similar behavior is also observed in the Heisenberg
SK model. The SG exponents in 5D are estimated as $\alpha=-1.0(5)$,
$\beta_{{\rm SG}}=0.7(3)$,
$\gamma_{{\rm SG}}=1.7(5)$ and $\nu_{{\rm SG}}=0.6(2)$,
most of which are rather close to
the mean-field exponents.
Since the upper critical dimension of the
SG is believed to be six, the observed closeness to the mean-field values
seems  reasonable.
Our data suggest that the SG ordered state
accompanies a peculiar
phase-space structure, namely, a one-step-like RSB, at least in its chiral
sector.
Such a one-step-like character of the RSB is at variance with
the full (hierarchical)
RSB realized in the SK model corresponding to $D=\infty$.
It means that the RSB pattern of
the ordered state changes its nature at some
borderline dimensionality, presumably at the upper critical dimension
$D=6$.

In 4D, the model exhibits a significantly different behavior from the
5D and the SK models.
Bulk of our data, particularly the Binder ratio,
indicate that the 4D model
exhibits a pure chiral-glass transition at a finite temperature
$T_{{\rm CG}}=0.38(2)$ without accompanying the standard SG order.
At the chiral-glass transition, however, the spin becomes almost
critical which manifests itself as a pseudo-critical phenomenon observable
at short length scales. The critical region associated with the chiral-glass
transition is very narrow, limited to $t\lsim 10^{-2}$,
suggesting that the dimension four
is close to the marginal dimensionality.
The SG transition occurs either at $T_{{\rm SG}}=0$
or at a finite temperature, but below the chiral-glass transition temperature,
$T_{{\rm SG}}<T_{{\rm CG}}$.
Our data suggest that the chiral-glass ordered state accompanies a
one-step-like RSB in the chiral sector. Again,
such a one-step-like character of the RSB differs from the full (hierarchical)
RSB realized in the SK model.

Next, we wish to compare out present results on the 4D and 5D Heisenberg SGs
with those of the previous authors. To our knowledge,
our results for the chiral order are new.
Concerning the spin order, our present conclusion, {\it i.e.\/}, the
presence of the SG LRO in 5D and the absence of it in 4D, is consistent
with most of numerical simulations, in particular, with the one
of Stauffer and Binder~\cite{Stauffer}.
Our conclusion, however, is at variance with that of Coluzzi, who suggested
that the SG LRO set in at a finite temperature $T_{{\rm SG}}/J\simeq 0.5$
~\cite{Coluzzi}.
Although the numerical
data themselves seem to be consistent between the two works,
Coluzzi simulated rather small lattices $L\leq 5$
and high temperatures $T/J\geq 0.5$,
which hampered a direct examination of the asymptotic ordering behavior.
Our new data for larger lattices $L\leq 10$ and for temperatures including
lower ones, $T/J\geq 0.26$, have clarified that the transition occurs
in the chiral sector at $T_{{\rm CG}}/J\simeq 0.4$, which is somewhat lower
than $T_{{\rm SG}}$ estimated in \reftag~\cite{Coluzzi}. Furthermore,
the Heisenberg spin appears to
remain paramagnetic at the chiral-glass transition point on
sufficiently long length scales, {\it i.e.\/}, the transition at
$T_{{\rm CG}}/J\simeq 0.4$ is {\it not\/}
the conventional SG transition, but a pure
chiral-glass transition.

Finally, we wish to discuss implication of our present result to
the 3D case. The behavior of the 4D model observed in the present work
is qualitatively similar to the one of the 3D model observed in
\refstag\cite{HK1,HK2}, except that the behavior of the
4D model looks much more marginal. For example,
the reduced chiral-glass susceptibility of the 3D model is much larger
in magnitude than that of the 4D model, and the associated
chiral-glass critical region is much wider in 3D than in 4D. As one judges
from the size dependence of the reduced chiral-glass susceptibility shown in
\figtag 3 of \reftag~\cite{HK1}, the width of the chiral critical region
is about $10^{-1}$, which
should be compared with our present estimate for the 4D model, $10^{-2}$.
All these suggest that the spin-chirality decoupling is more eminent in lower
dimensions. As the dimensionality is increased, the spin-chirality
decoupling tends to be suppressed. In 4D, the spin-chirality decoupling still
seems to persist, but it is limited only to a very narrow temperature region
close to the transition temperature, suggesting that
4D is close to the borderline dimensionality. As the dimensionality is
further increased, the spin-chirality decoupling no longer occurs. There, the
order parameter of transition is the spin, not the chirality.
This is indeed the case for 5D.
However, at least in the case of 5D, the SG ordered
state exhibits a peculiar one-step-like RSB,
which differs in character
from the full RSB of the $D=\infty$ SK model.
Estimated SG critical exponents of the 5D model are rather close to the
mean-field values, which is consistent with a common belief that the
mean-field SG exponents arise above six dimensions.

\section*{Acknowledgements}
The numerical calculation was performed on the HITACHI SR8000
at the supercomputer system, ISSP, University of Tokyo,
and Intel Pentium4 1.8GHz PCs in our laboratory.
The authors are
thankful to Dr. K. Hukushima and Dr. H. Yoshino for useful discussion.

%
\appendix
\section{Derivation of \eqtag(\ref{eqn:Pdform})}
\label{devPdform}
\setcounter{equation}{0}
In this appendix,
we give the derivation of \eqtag(\ref{eqn:Pdform}). It describes
the self-overlap part of the diagonal spin-overlap distribution function
in the thermodynamic limit, when
the SG ordered state with a nonzero EA order parameter
$q_{\rm s}^{{\rm EA}}>0$
exists.
Since the diagonal spin-overlap $q_{\rm diag}$ transforms nontrivially
under global $O(3)$ rotations, even a self-overlap part of the
distribution function is not just a simple delta function
located at $q_{\rm diag}=\pm q_{\rm s}^{{\rm EA}}$,
but exhibits a nontrivial behavior given by \eqtag(\ref{eqn:Pdform}).

\begin{figure}[ht]
\begin{center}
\psfrag{xlabel}{$x=\cos\Theta$}
\psfrag{ylabel}{$q_{\rm diag}(x,\alpha)$}
\psfrag{origin}{O}
\psfrag{xmax}{$1$}
\psfrag{xmin}{$-1$}
\psfrag{ymax}{$1$}
\psfrag{ymin}{$-1/3$}
\psfrag{ycut}{$1/3$}
\psfrag{alpha1}{$\alpha=0$}
\psfrag{alpha2}{$\alpha=\alpha_c$}
\psfrag{alpha3}{$\alpha=\pi/2$}
\psfrag{alpha4}{$\alpha=\pi$}
\includegraphics{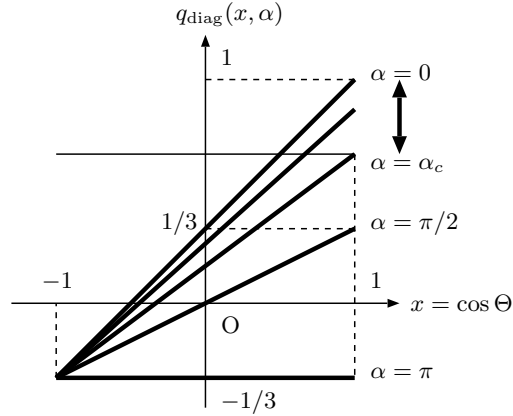}
\caption{Sketch of $q_{\rm diag}$ as a function of $x$
for various $\alpha$. For a given value of $q_{{\rm diag}}$,
there is
no possible value of $x$ for a range of $\alpha$ satisfying
$\alpha_c(q_{{\rm diag}})\leq \alpha \leq \pi$.
}
\label{fig_q-x}
\end{center}
\end{figure}

We consider a diagonal spin-overlap between a particular spin
state described by the
configuration $\vec{S}_i$ and a state
generated from this state via a global $O(3)$ rotation $R$,
\begin{equation}
q_{\rm diag}=\frac{1}{N}\sum_{i=1}^N \vec{S}_i\cdot R\vec{S}_i\ \ .
\label{eqn:qdappndx}
\end{equation}
We first consider the case of proper rotations with det($R$)=1.
The $SO(3)$ rotation matrix $R$ is known to be
parametrized by the Euler angles, $\Phi$,
$\Theta$ and $\Psi$, as
\begin{center}
\[
R=
\left(
\begin{array}{ccc}
R_{xx} & R_{xy} & R_{xz} \\
R_{yx} & R_{yy} & R_{yz} \\
R_{zx} & R_{zy} & R_{zz} \\
\end{array}
\right)
=
\left(
\begin{array}{ccc}
\cos\Theta\cos\Phi\cos\Psi-\sin\Phi\sin\Psi &
\cos\Theta\sin\Phi\cos\Psi+\sin\Phi\cos\Psi &
-\sin\Theta\cos\Psi \\
-\cos\Theta\cos\Phi\sin\Psi-\sin\Phi\cos\Psi &
-\cos\Theta\sin\Phi\sin\Psi+\cos\Phi\cos\Psi &
\sin\Theta\sin\Psi \\
\sin\Theta\cos\Phi & \sin\Theta\sin\Phi & \cos\Theta \\
\end{array}
\right)\ \ .
\]
\end{center}
Then, $q_{\rm diag}$ can be written as
\begin{equation}
q_{\rm diag}=\frac{1}{N}\sum_{i=1}^N(
R_{xx}S_{ix}^2+R_{yy}S_{iy}^2+R_{zz}S_{iz}^2
+(R_{xy}+R_{yx})S_{ix}S_{iy}
+(R_{yz}+R_{zy})S_{iy}S_{iz}
+(R_{zx}+R_{xz})S_{iz}S_{ix}
)\ \ .
\label{eqn:q_with_R}
\end{equation}
The spin direction at each site can be represented
as $\vec{S}_i=(\sin\theta_i\cos\phi_i,
\sin\theta_i\sin\phi_i,\cos\phi_i)$.
If one notes the fact that, in the SG ordered state, the spin direction
is entirely random on long length scales giving
a uniform distribution on a sphere in  spin space, one can
replace in the thermodynamic limit the summation over spins
by the integral over spin directions as
$(1/N)\sum_{i=1}^N \rightarrow(1/4\pi)\int_{-1}^1{\rm d}\cos\theta
\int_0^{2\pi}{\rm d}\phi$.
Then, only the diagonal terms containing
$R_{\mu\mu}$ survive in \eqtag(\ref{eqn:q_with_R}),
leading to
\begin{equation}
q_{\rm diag}=\frac{\cos(\Phi+\Psi)+1}{3}\cos\Theta
+\frac{\cos(\Phi+\Psi)}{3}
=\frac{\cos(\Phi+\Psi)+1}{3}x
+\frac{\cos(\Phi+\Psi)}{3}\ \ ,
\label{eqn:qRel}
\end{equation}
where $x\equiv \cos\Theta$. Note that this is a function of the rotation matrix
$R$ only, not depending on the spin configuration $\vec S_i$ any more.
The overlap $q_{{\rm diag}}$
takes various values depending on the $O(3)$ matrix $R$.
We then consider the distribution of $q_{\rm diag}$ arising from the
distribution of $R$, or equivalently, $x$, $\Phi$ and $\Psi$.
The appropriate measure is $-1\leq x\leq 1$,
$0\leq\Phi<2\pi$ and $0\leq\Psi< 2\pi$ being uniform.
It is convenient to change the variables from $(\Phi, \Psi)$ to
$(\alpha, \beta)=(\Phi+\Psi, (-\Phi+\Psi)/2)$,
where $0\leq\alpha<4\pi$ and $0\leq\beta<\pi$.
With this change of the
variables, \eqtag(\ref{eqn:qRel}) becomes independent of
$\beta$, and is given by
\begin{equation}
q_{\rm diag}=\frac{\cos\alpha+1}{3}x+\frac{\cos\alpha}{3}\ \ .
\end{equation}
The distribution function $P_{\rm s}(q_{\rm diag})$ is
proportional to
\begin{eqnarray}
P_{\rm s}(q_{\rm diag})
&\propto&
\int\frac{{\rm d}x}{{\rm d}q_{\rm diag}}{\rm d}\alpha
\nonumber\\
&=&\int_0^{\alpha_c(q_{\rm diag})}\frac{3}{1+\cos\alpha}
{\rm d}\alpha\ \ ,
\label{eqn:PqResult}
\end{eqnarray}
where we have used \eqtag(\ref{eqn:qRel}).
Note that, for a given $q_{\rm diag}$,
the integral with respect to $\alpha$ is
restricted to the range $[0, \alpha_c]$, with
$\alpha_c(q_{\rm diag})=\cos^{-1}[(3q_{\rm diag}-1)/2]$.
This may be seen from
\figtag\ref{fig_q-x}, where we plot $q_{\rm diag}$ as a function of
$x$ for various $\alpha$. Obviously, for a given $q_{\rm diag}$,
no contribution to the integral arises from the region of $\alpha$
between [$\alpha_c, \pi$]. The  $q_{\rm diag}$-dependence of
$P_{\rm s}(q_{\rm diag})$ arises from this upper limit of the integral.
The integration in \eqtag(\ref{eqn:PqResult}) can be easily carried out
to yield,
\begin{displaymath}
P_{\rm s}(q_{\rm diag})\propto
\tan\frac{\alpha_c(q_{\rm diag})}{2}
=3^{1/2}\sqrt{\frac{1-q_{\rm diag}}{3q_{\rm diag}+1}}\ \ .
\end{displaymath}

So far, we have considered proper rotations.
The contribution
from improper rotations, which can be viewed as
proper rotations combined with the spin inversion
$\vec{S}_i\rightarrow -\vec{S}_i$,
may be obtained  immediately by the replacement
$q_{\rm diag}\rightarrow -q_{\rm diag}$.
Adding the contributions from both proper and improper rotations
with equal weights, and reproducing the appropriate normalization
factor, we get
\begin{equation}
P_{\rm s}(q_{{\rm diag}})=
\frac{3\sqrt{3}}{4\pi}\left(
\sqrt{\frac{1-q_{\rm diag}}{3q_{\rm diag}+1}}
+\sqrt{\frac{1+q_{\rm diag}}{-3q_{\rm diag}+1}}
\right)\ \ .
\label{eqn:PdLastForm}
\end{equation}
Finally, we note that at finite temperatures a state
should be regarded as a pure state. The spin length is then no longer
unity, and the unity in \eqtag(\ref{eqn:PdLastForm}) should be
replaced by $q_{\rm s}^{\rm EA}$. We finally obtain
\begin{equation}
P_{\rm s}(q_{{\rm diag}})=
\frac{3\sqrt{3}}{4\pi q_{\rm s}^{\rm EA}}\left(
\sqrt{
\frac{q_{\rm s}^{\rm EA}-q_{\rm diag}}{3q_{\rm diag}+q_{\rm s}^{\rm EA}}}
+\sqrt{
\frac{q_{\rm s}^{\rm EA}+q_{\rm diag}}{-3q_{\rm diag}+q_{\rm s}^{\rm EA}}}
\right)\ \ ,
\label{eqn:PdLastForm2}
\end{equation}
which is \eqtag(\ref{eqn:Pdform}).

\begin{figure}[ht]
\begin{center}
\includegraphics{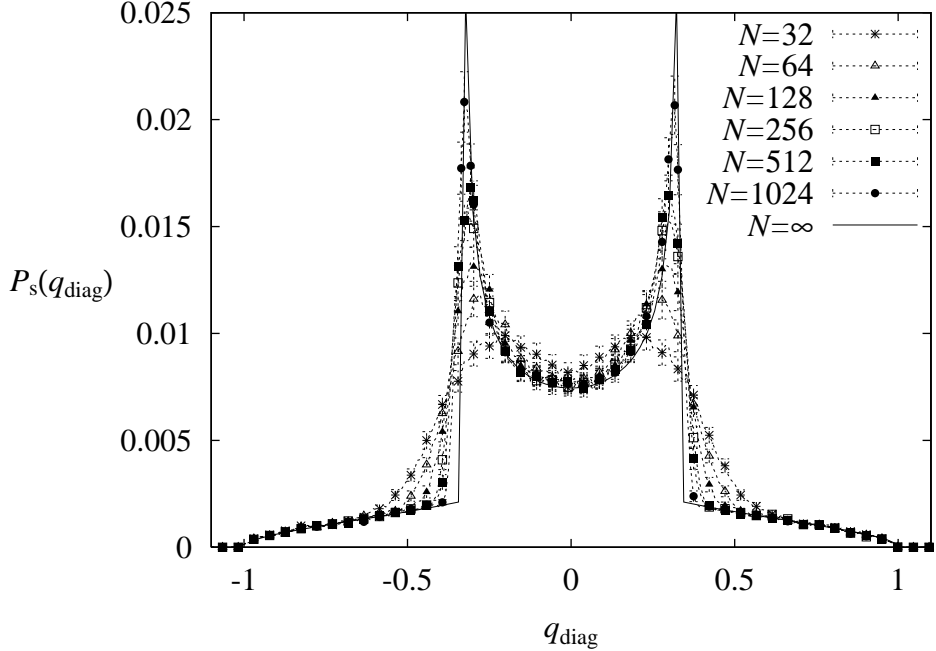}
\caption{The diagonal spin-overlap distribution function
$P_{\rm s}(q_{\rm diag})$ of finite-size systems of
$N$ Heisenberg spins with completely random, uncorrelated spin configurations.
The result for $N=\infty $ given by \eqtag(\ref{eqn:Pdform})
is also shown by the solid curve. For further details, see the text.}
\label{fig_PdFS}
\end{center}
\end{figure}

The derivation above, valid in the thermodynamic limit $N\rightarrow\infty$,
is quite general.
In order to get some feeling about the finite-size effect,
we also compute $P_{\rm s}(q_{\rm diag})$ numerically for finite-$N$
Heisenberg spins,
the direction of which is assumed to be completely random in three-component
spin space. More specifically, we prepare a random and uncorrelated
configuration of $N$ spins,
numerically generate $O(3)$ rotation matrix $R$
with appropriate measure ({\it i.e.\/},
the one generated randomly from the uniform distribution of
$-1\leq x\leq 1$, $0\leq \Phi<2\pi$, $0\leq \Psi<2\pi$ and
the determinant $\pm 1$), operate $R$ to the initial spin configuration,
and compute the diagonal spin-overlap $q_{\rm diag}$ between the initial
and the $O(3)$-rotated spin configurations.
We generate $10^4$ distinct $O(3)$ matrices for a given initial spin
configuration, and generate several hundreds of initial
spin configurations, $P_{\rm s}(q_{\rm diag})$  being accumulated over
these procedures.
The result is shown in \figtag\ref{fig_PdFS}.
The  $N=\infty $ result  analytically obtained above
is also shown.
It can be seen that the rounded peak at
$\pm \frac{1}{3}q_{\rm s}^{\rm EA}$ grows as $N$ increases,
eventually exhibiting a divergent behavior in the $N=\infty $ limit.
Of course, the finite-$N$ result computed here is valid
only for non-interacting
spins. It would differ from the corresponding result for the interacting
system, in contrast to the analytical $N=\infty $ result which is valid
even for the interacting system.
It still gives some feeling how the finite-size rounding takes place
in finite-$N$ SG models.
%
%


\end{document}